\documentclass[twocolumn,prd,aps,superscriptaddress,preprintnumbers,tightenlines,showpacs,nofootinbib,eqsecnum,amsfonts,amsmath]{revtex4-1}
\usepackage{placeins}
\usepackage{color}
\usepackage{calc}
\usepackage{amsmath,amssymb,graphicx,mathrsfs}
\usepackage{tensor}
\usepackage{bm}
\usepackage{times}
\usepackage[varg]{txfonts}
\usepackage[colorlinks, pdfborder={0 0 0}]{hyperref}
\usepackage{float}
\usepackage{physics}
\usepackage[nolist,nohyperlinks]{acronym}
\usepackage{xspace}
\usepackage[abs]{overpic}
\usepackage{pict2e}
\usepackage{enumitem}
\usepackage[usenames,dvipsnames]{xcolor}
\usepackage[utf8]{inputenc}
\usepackage{acronym}
\usepackage{gensymb}
\usepackage{romannum}
\usepackage{subfigure}
\usepackage{cleveref} 
\usepackage[normalem]{ulem}
\usepackage{longtable}
\usepackage{mathtools}
\usepackage{multirow}
\usepackage{dcolumn}

\usepackage[normalem]{ulem}

\xdefinecolor{mylinkcolor}{rgb}{0,0,0.5}

\allowdisplaybreaks

\newcommand{\AEI}{\affiliation{Max Planck Institute for Gravitational Physics (Albert Einstein Institute), Am M\"uhlenberg 1, Potsdam 14476, Germany}}
\newcommand{\Maryland}{\affiliation{Department of Physics, University of Maryland, College Park, MD 20742, USA}}
\newcommand{\Stanford}{\affiliation{Department of Physics, Stanford University, Stanford, CA 94306, USA}}
\newcommand{\Sapienza}{\affiliation{Dipartimento di Fisica, ``Sapienza'' Universit\`a di Roma \& Sezione INFN Roma-1, Piazzale Aldo Moro 5, 00185, Roma, Italy}}
\newcommand{\Princeton}{\affiliation{School of Natural Sciences, Institute for Advanced Study, Princeton, NJ 08540, USA}}
\newcommand{\SLAC}{\affiliation{Kavli Institute for Particle Astrophysics and Cosmology, SLAC and Stanford University, Menlo Park, CA 94025, USA}}

\begin{document}

\pagenumbering{arabic} 

\title{Gravitational-Wave Constraints on an Effective--Field-Theory Extension of General Relativity}
\author{Noah Sennett} \email{noah.sennett@aei.mpg.de} \AEI \Maryland
\author{Richard Brito} \Sapienza \AEI
\author{Alessandra Buonanno} \AEI \Maryland 
\author{Victor Gorbenko} \Princeton\Stanford
\author{Leonardo Senatore} \Stanford\SLAC
\date{\today}

\begin{abstract}
  Gravitational-wave observations of coalescing binary systems allow
  for novel tests of the strong-field regime of gravity.  Using data
  from the Gravitational Wave Open Science Center (GWOSC) of the LIGO
  and Virgo detectors, we place the first constraints on an
  effective--field-theory based extension of General Relativity in
  which only higher-order curvature terms are added to the Einstein-Hilbert action. We construct gravitational-wave templates describing the
  quasi-circular, adiabatic inspiral phase of binary black holes in this extended theory of
  gravity. Then, after explaining how to properly take into account the region of validity of the 
effective field theory when performing tests of General Relativity, 
we perform Bayesian model selection using the two
  lowest-mass binary--black-hole events reported to date by LIGO and Virgo ---GW151226
  and GW170608--- and constrain this theory with respect to General
  Relativity. 
  We find that these data disfavors the appearance of
  new physics on distance scales around $\sim 150$ km. Finally, we describe a general strategy for improving constraints as more observations will
  become available with future detectors on the ground and in space.
\end{abstract}

\maketitle

\section{Introduction}

The detections of gravitational waves (GWs) from merging black-hole (BH) and neutron-star binaries by the LIGO and Virgo collaborations~\cite{LIGOScientific:2018mvr} provide a novel opportunity to test the highly-dynamical, strong-field regime of gravity. All tests to date indicate that these observations are fully consistent with the predictions of General Relativity (GR)~\cite{TheLIGOScientific:2016src,Yunes:2016jcc,Abbott:2018lct,LIGOScientific:2019fpa}. Combined with tests from the Solar System and cosmological observations~\cite{Will:2014kxa,Baker:2014zba,Baker:2019gxo}, we now have a great amount of evidence that GR accurately describes gravity over a broad range of scales.

While positive checks for consistency are significant results by
themselves, it remains important also to consider how GW observations
can directly inform our understanding of how gravity behaves in the
high-curvature regime. Work towards this second goal typically
proceeds in one of two directions. One option is to consider a
particular alternative to GR (typically at the level of an action),
then calculate how the differences in the underlying physical theory
translate to an observable signal, i.e., the gravitational waveform,
and finally use a detected GW to measure the physical parameters that
define that theory. While straightforward, this approach suffers due
to its specificity; a plethora of proposals for modifying GR in the
high-curvature regime have been considered~\cite{Berti:2015itd}, and
testing each individually is highly inefficient. The second option
instead considers phenomenological
deviations~\cite{Arun:2006hn,Yunes:2009ke,Mishra:2010tp,Li:2011cg,Agathos:2013upa}
from the expected GW signal in GR, it uses observations to constrain
these deviations, and then maps those bounds to constraints on
specific non-GR theories~\cite{Yunes:2016jcc,Nair:2019iur}. But, this
approach does not always provide a clear connection to the fundamental
physics/principles one wishes to test.

In this work, we adopt an alternative approach, employing the powerful
tools of Effective Field Theory (EFT) to test natural extensions of GR
with GW observations.  Broadly speaking, EFT provides a systematic
framework to encode all modifications to an existing theory that could
arise after introducing some form of \textit{new physics}.  Though this
task may seem daunting, the construction of an EFT is dramatically
simplified by restricting ourselves to modifications that: (i) respect
certain general principles or symmetries (e.g., locality and Lorentz
invariance), and (ii) are associated with a particle (or some form of
undetermined excitation) whose mass is too large to be directly probed
with experiments. Under these assumptions, the Lagrangian of our EFT
is determined by simply adding to the original Lagrangian all possible
terms allowed by the symmetries that we wish to preserve, constructed
using the fields already present in the original theory. Dimensional
analysis dictates that these new terms are each suppressed by some
energy scale, dubbed collectively as the \textit{cutoff scale} $\Lambda_c$,
which, roughly speaking, corresponds to the mass of the particles that
modify the original theory. Note that describing physics at energies
above this cutoff would require the introduction of new particles to
the theory, or equivalently, new fields to the effective Lagrangian,
vastly complicating the problem\footnote{Specific examples of theories with additional degrees of freedom, in this case a scalar field, include dynamical Gauss-Bonnet and dynamical Chern-Simons.}. However, provided we work at energies
below the cutoff scale [assumption (ii) above], all physical
observables can be estimated without specifying these new particles in
detail. The high-energy-agnosticism of the EFT provides the primary
advantage of this approach over the tests of specific alternative
theories described earlier; by working within the confines of an EFT,
we can simultaneously test a large range of possible extensions of GR
at once.  
 
Reference~\cite{Endlich:2017tqa} constructed an EFT suitable for testing gravity with current GW experiments following the guidelines above. 
In this paper, we examine the constraints that can be set on this extension of GR given existing GW observations. 
The work is organized as follows. In Sec.~\ref{sec:EFT}, we review the main properties of the EFT extension of GR constructed in 
Ref.~\cite{Endlich:2017tqa}. In Sec.~\ref{sec:waves} we construct gravitational waveforms that describe the quasi-circular, 
adiabatic inspiral of BBHs in this theory. Then, in 
Sec.~\ref{sec:results}, we employ Bayesian inference, notably the model-selection method, to place constraints on this theory when 
$\Lambda_c \lesssim 1/r_s$, $r_s$ being the BH radius, 
demonstrating that current GW data rule out modifications to GR entering at scales around $\Lambda_c^{-1} \sim 150$ km (or equivalently, $\Lambda_c \sim 1.3$ peV). 
Finally, we provide some concluding remarks and discuss future directions in Sec.~\ref{sec:conc}. Lastly, in 
the Appendix~\ref{app:finitesize}, we consider setting constraints in the regime $\Lambda_c\gtrsim 1/r_s$, 
while in the Appendix~\ref{app:IMR} we discuss how our findings 
are modified if an inspiral-merger-ringdown waveform model instead of an inspiral-only model were used.  
Henceforth, we use natural units $G = c=  1$ except when stated otherwise.

\section{Effective--field-theory extension of General Relativity}\label{sec:EFT}

\subsection{Extension of General Relativity using high powers in Riemann tensor}

The EFT constructed in Ref.~\cite{Endlich:2017tqa} encodes the most
general extension to gravity under a number of assumptions: (i) the
following principles and symmetries are preserved by the new physics:
locality, causality, Lorentz invariance, unitarity and diffeomorphism
invariance, (ii) there is no new particle lighter than the cutoff of
the theory, and (iii) the extension to the theory is testable with
  experiments such as LIGO and Virgo.  The resulting EFT takes the
form
\begin{equation}\label{EFT_action}
S_{\rm eff}=2M_{\rm pl}^2\int d^4 x \sqrt{-g} \left(-R+\frac{\mathcal{C}^2}{\Lambda^6}+\frac{\tilde{\mathcal{C}}^2}{\tilde{\Lambda}^6}+\frac{\mathcal{C}\,\tilde{\mathcal{C}}}{\Lambda_-^6}\right) + \ldots\,,
\end{equation}
where $M_\text{pl}=\sqrt{\hbar}$ is the Planck mass, %
\begin{equation}
\mathcal{C}\equiv R_{\alpha\beta\gamma\delta}\,R^{\alpha\beta\gamma\delta}\,, \quad
\tilde{\mathcal{C}}\equiv R_{\alpha\beta\gamma\delta}\,\epsilon^{\alpha\beta\phantom{\mu}\phantom{\nu}}_{\phantom{\alpha}\phantom{\beta}\mu\nu}\,R^{\mu\nu\gamma\delta}\,,  \label{eq:Cdef}
\end{equation}
where $\Lambda, \tilde{\Lambda}, \Lambda_-$ are cutoff scales for each
  operator, and the dots in Eq.~(\ref{EFT_action}) denote terms with
powers in the Riemann tensor beyond four. Though the cutoff scale
  for each term is, in principle, independent, it is quite natural to
  assume that they are comparable (i.e., $\Lambda \sim
  \tilde{\Lambda}\sim \Lambda_-$.) In Eq.~\eqref{eq:Cdef}, we
  employ the Levi-Civita tensor $\epsilon^{\mu \nu \rho \sigma}$,
  defined such that $\epsilon^{0123}=1/\sqrt{-g}$. We refer to
  the EFT~\eqref{EFT_action} as ``EFT of General Relativity'', in short {\tt EFTGR}.
Except where noted, we assume the coupling constants to be positive 
(i.e., $\Lambda^6,\tilde{\Lambda}^6,\Lambda_-^6>0$), however, in principle
  one should not dismiss the parameter space $\Lambda^6,
  \tilde{\Lambda}^6,\Lambda_-^6<0$~\cite{Endlich:2017tqa}~\footnote{It
  has been argued that causality~\cite{Gruzinov:2006ie} and
  analyticity of the graviton amplitudes~\cite{Bellazzini:2015cra}
  requires $\Lambda^6,\tilde{\Lambda}^6>0$ and $\Lambda_-^{12}\leq
  2/(\Lambda^6\tilde{\Lambda}^6)$~\cite{Endlich:2017tqa}. However,
  these arguments seem to require the graviton's momenta to be larger
  that the cutoff scale $\Lambda$ and therefore, strictly speaking, to
  be outside of the regime of validity of the
  theory~\cite{Endlich:2017tqa}. We therefore regard these arguments
  just as highly disfavoring these theories.}.
  
It is well known that the terms in Eq.~\eqref{EFT_action}, as well as
  others, are needed by the renormalizability of the theory (e.g., 
  see Ref.~\cite{Goroff:1985th}). Renormalizability, however,
  suggests that the scales suppressing
  those operators are different from the ones in Eq.~\eqref{EFT_action}. 
  In fact, the
  main novelty of Eq.~\eqref{EFT_action} lies in realizing that
  such low-energy suppression for those operators and such scalings
  were possible from an EFT point of view, and, as we will describe,
  are not yet ruled out by tests of gravity. The label of {\tt EFTGR}
  is referred in this paper only to the particular scaling assumed
  in Eq.~\eqref{EFT_action}, which gives a theory testable by LIGO and Virgo, 
  and not  to the general, and already well-established fact, that GR is 
  an EFT (e.g., see Ref.~\cite{Endlich:2017tqa} for detailed
  discussions).

For clarity, we now briefly review the construction of the action
(\ref{EFT_action}); the interested reader should refer to
Ref.~\cite{Endlich:2017tqa} for more detail.  As discussed above,
  our assumption that no new particle enters with a mass below the
  cutoff of the EFT precludes the introduction of any new fields to
  the theory; thus our effective action must only be a function of the
  metric $g_{\mu \nu}$.  Furthermore, our assumption of
  diffeomorphism invariance implies that new terms must be
  expressible as functions of the Riemann tensor and its covariant
  derivatives.  Consider a Taylor expansion of such a function:
dimensional analysis dictates that terms polynomial in the Riemann
tensor must be suppressed by powers of some cutoff scale $\Lambda_c$. 
In principle, we could write an infinite number of higher-order curvature terms 
suppressed by correspondingly large powers of some cutoff scale 
[i.e., terms given schematically by $(R_{\mu\nu\rho\sigma}/\Lambda_c^2)^n$]. 
But, if we are interested in observables whose energy scale $E$ is much 
smaller than $\Lambda_c$ (as in the present context), then each term, schematically 
of the form $(R_{\mu\nu\rho\sigma}/\Lambda^2)^n$, contributes as $(E^2/\Lambda^2)^n\ll 1$. 
Therefore, for a given precision of the prediction, only a finite number of terms needs to be kept.

For our purposes, we keep only the leading-order corrections to the Einstein-Hilbert action; 
these are precisely the terms given in Eq.~\eqref{EFT_action}. However, our truncation of the 
higher-order curvature corrections remains valid only so long as the sub-dominant terms remain 
negligible at the level of precision at which we work. Yet, working within this restricted regime 
of validity, we see that any extension of GR that upholds conditions (i) and (ii) described 
above must admit a low-energy~\footnote{Here by energy we mean the magnitude of the gradients, not 
the total energy present in the system.} EFT of the form~\eqref{EFT_action}.

Note that the leading-order corrections to the Einstein-Hilbert
Lagrangian come from terms that are quartic in the Riemann tensor (or
equivalently, eight derivatives of the metric) rather than terms
quadratic or cubic in curvature. In vacuum (to which we restrict our
attention), terms quadratic in the Riemann tensor can be removed via
an appropriate redefinition of the gravitational field~\footnote{In
  presence of matter, this field redefinition introduces a
  matter-matter interaction of the form $TT/(M_{\rm pl}^2
  \Lambda_c^2)$, where $T T$ is some scalar term built out of two
  matter-stress tensors. For the scales at which we are interested in,
  this interaction is highly ruled out. One can take two points of
  view to interpret these terms. Either they represent purely matter
  interactions, and so one does not need to include them to start
  with, as they do not appear to be related to extensions to
  gravity. Alternatively, even if they are included among the terms in
  the Lagrangian, they are experimentally ruled out, and then one
  moves on to the next testable theory, as we do.}.  Our dismissal of
terms cubic in Riemann is more subtle. There are several theoretical
arguments that disfavor the presence of these terms, see
Refs.~\cite{Camanho:2014apa,Reall:2019sah}, as well as Sec. 2.2 of
Ref.~\cite{Endlich:2017tqa}.  Additionally, there are several
  cancellations in the post-Newtonian (PN) calculation of the contribution of these
  terms to the GW signal \cite{Endlich:2017tqa}. This leads to the
  suppression of these effects, going as $v^4$ instead of $v^2$, as
  instead expected from naive estimates, and makes the computation of
  the non-vanishing effects a harder task due to proliferation of
  Feynman diagrams, as typically occurs once one goes beyond the
  leading order.
 In spite of these facts, further study of
these terms remains an interesting direction that we leave for future
work.

Finally, one can, in principle, include in the action~\eqref{EFT_action} 
additional terms containing covariant derivatives of the Riemann tensor 
that equate to eight derivatives of the metric; however, these terms can 
always be absorbed into the existing terms in Eq.~\eqref{EFT_action} 
via integration by parts (see Sec.~2.1 of Ref.~\cite{Endlich:2017tqa}), and
so we can work with the simplified action without any loss of generality.

Clearly, the higher-order curvature terms in Eq.~(\ref{EFT_action}) produce  
the greatest deviation from GR when the energy scales in the problem approach 
the cutoff scale of the EFT. In the context of a BBH system, this occurs when 
the cutoff scale is comparable to the inverse Schwarzschild radius of the merging BHs 
(the shortest curvature scale in the problem). For typical stellar-mass BHs observable 
by the LIGO and Virgo detectors, this corresponds to an energy scale of the order of 
inverse kilometers (or tens of picoelectron volts).

By  particle physics standards, this energy cutoff is quite low. Indeed, there is no 
\textit{a priori} expectation that GR should receive corrections at such low-energy scales. 
Yet, there remain several factors that merit further exploration of this EFT despite its contrivance. 
First, as we will discuss in more detail below, under some mild assumptions on the ultraviolet (UV) 
completion of this EFT, there is no experimental evidence that gravity excludes the possibility of new physics 
on the scale of inverse kilometers. Second, a presupposition of naturalness in our EFT is itself a theoretical bias; 
given the wealth of observational data now available from the LIGO and Virgo experiments, it is worthwhile to adopt 
a more agnostic viewpoint, letting the data determine the correct description of nature rather than our theoretical 
preconceptions. Finally, the Lagrangian in Eq.~(\ref{EFT_action}) with cutoff  $\Lambda_c$ is consistent at classical 
and quantum level even for very low-energy cutoff $\Lambda_c$.

\subsection{Soft ultraviolet  completion of the effective--field-theory extension of General Relativity}

The predictions of GR have been tested with exquisite precision in
laboratory on the Earth and Solar-System
experiments~\cite{Will:2014kxa}.  Naively, one may expect that
such experiments could be sensitive to modifications of GR that enter
at scales of inverse kilometers in the form of
Eq.~\eqref{EFT_action}.  However, there exist conditions under
which the extensions of GR that we consider completely circumvent
these weak-field constraints while still impacting strong-field
phenomena detectable through GWs from stellar-mass BHs. In this subsection, we outline
this additional condition precisely, which we denote as the existence
of a \textit{soft ultraviolet (UV) completion}.

The key distinction between the weak-field and strong-field regimes alluded 
to above is not the distance scales of the systems involved, but rather the 
curvature scales manifested in each. Typical curvature scales in the Solar System 
and terrestrial experiments are of the order of less than $10^{-8}\text{ km}^{-1}$~\cite{Baker:2014zba}, 
whereas stellar-mass BHs generate curvatures up to scales of inverse kilometers.
Under the assumption of a soft UV completion, deviations from GR like those in Eq.~\eqref{EFT_action} 
can still arise at cutoff scales $\Lambda_c$ of inverse kilometers, and yet have negligible effects 
on weak-field tests. A different point of contention may arise from tests of gravity that involve 
$X$-ray binaries~\cite{Bambi:2014nta,Reynolds:2013qqa}; however, the precision of current observations 
is not high enough to rule out the models we consider here (see Sec.~8 of Ref.~\cite{Endlich:2017tqa} 
for further discussion of all relevant experiments).

To outline what is meant by a soft UV completion, we first recall that the EFT in Eq.~\eqref{EFT_action} 
breaks down at the scale $\Lambda_c$. Thus, to make predictions on shorter distances scales, one needs to 
introduce new degrees of freedom and use a more fundamental theory. As a result, one cannot guarantee that an 
EFT of the form~\eqref{EFT_action} reduces to GR on distance scales shorter than $\Lambda_c^{-1}$. 
At energies (or inverse distances) below $\Lambda_c$, all EFT effects grow as $(E/\Lambda_c)^n$ with $n>0$; 
if one extrapolates this growth to $E\gg\Lambda_c$, then the EFT effects grow very large at short distances. 
The assumption of a soft UV completion states exactly that this blow up at high energies does not occur.
Namely, we assume that all EFT effects saturate at the EFT breakdown point $E\sim \Lambda_c$ and either stay constant or even decay at higher energies.   

A similar sort of soft UV completion occurs in many familiar theories that have already been experimentally verified, most notably the Standard Model 
of particle physics. There, the scale $\Lambda_c$ is nothing but a mass of some heavy particle (e.g., the Higgs boson or the $W$-boson), $\Lambda_c\simeq m$. 
In these cases, at energies $E$ below the cutoff, the mass enters the predictions for many physical observables as $(E/m)^n$ with $n>0$; however, 
when $E>m$, the mass no longer enters in the denominator of any terms. Unfortunately, we are not able to provide any explicit example of a UV completion 
for our EFT (whether we require the softness assumption or not). But this fact only stresses the versatility of the EFT approach---we can make 
explicit predictions without the knowledge of the UV physics, which can be extremely complex. By working only within its regime of validity, we are 
able to place bounds on {\tt EFTGR} with GW observations with no further assumption about the underlying short-distance theory besides that of a soft UV completion.

Let us mention that the soft UV completion assumption appears to be a very generic necessary requirement for extensions of GR to be potentially testable 
by GW observatories --- for example there exist simple extensions of our EFT that differ by the addition of a single light scalar or pseudoscalar field which 
couples to the curvature invariant $\mathcal{C}$ or $\tilde{\mathcal{C}}$ correspondingly. These theories are usually called dynamical Gauss-Bonnet and 
dynamical Chern-Simons gravities in the literature (see, e.g., Ref.~\cite{Berti:2015itd}). Simple EFT analysis reveals that for the values of parameters 
testable by GW observatories these theories also must have a cutoff at the scale of order of the inverse Schwarzschild radius and require the same 
assumption about the short-distance behavior as our EFT~\cite{Nair:2019iur}. These features, required by theoretical consistency, are often missed 
since these theories are not properly treated as EFTs (see, however, the recent works~\cite{Okounkova:2017yby,Witek:2018dmd,Okounkova:2019dfo,Nair:2019iur}).
Possible infrared modifications of gravity, like theories of massive gravity, also require a similar sort of assumption about the UV completion~
\cite{Vainshtein:1972sx}. 

\subsection{Phenomenological consequences of the effective--field-theory extension of General Relativity}

\begin{figure*}[htb]
\begin{center}
\includegraphics[width=0.8\textwidth,clip=true,trim=0 10 0 0]{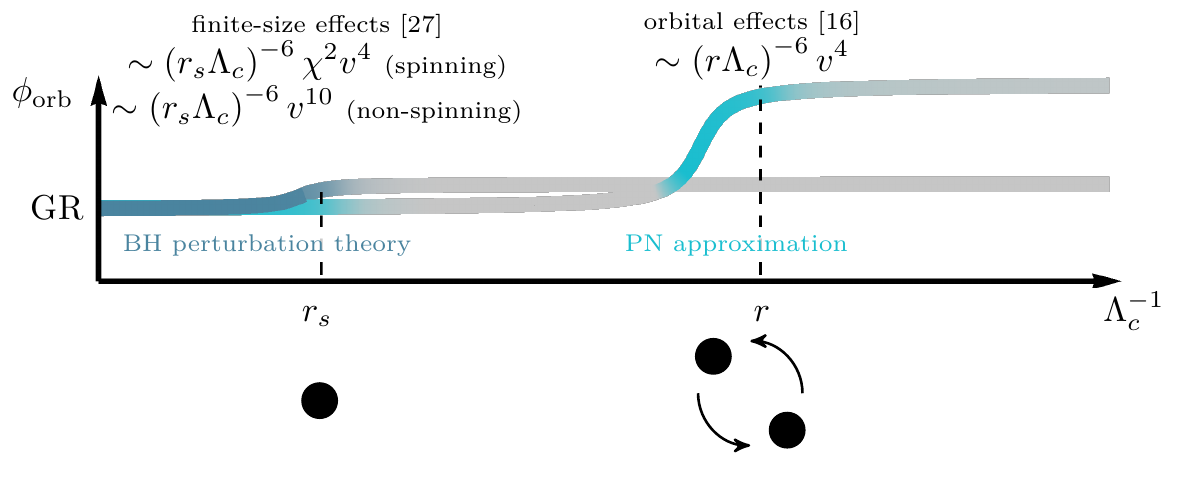}
\caption{The phenomenological consequences of \texttt{EFTGR} in binary BH systems and the regimes of applicability of perturbative approximations in which they are computed.
  The vertical axis schematically depicts the corrections to a generic observable (here taken to be the orbital phase $\phi_\text{orb}$) relative to its prediction in GR as function of the inverse cutoff scale $\Lambda_c^{-1}$.
  For extensions to GR at distance scales below the Schwarzschild radii of the BHs $r_s$, one can use BH perturbation theory to compute the finite-size effects that influence the binary dynamics~\cite{Cardoso:2018ptl}, shown in dark blue. For $\Lambda_c^{-1} \gtrsim r_s$, this approximation scheme breaks down, shown in gray, and the finite-size effects cannot be computed explicitly; however, the assumption of a soft UV completion ensures that these effects saturate at $\Lambda_c \sim 1/r_s$, changing by no more than a factor of order one for cutoff energies below this value.
  Similarly, new orbital effects~\cite{Endlich:2017tqa} are computed using the PN approximation; these corrections are subdominant to the aforementioned finite-size effects when the cutoff distance scale $\Lambda_c^{-1}$ is much smaller than the orbital separation $r$, but dominate when $\Lambda_c \sim 1/r$.
  However, by construction, $\texttt{EFTGR}$ is only valid over distance scales larger than the $\Lambda_c^{-1}$, and so the PN prediction of new orbital effects cannot be extended to separations below $\Lambda_c^{-1}$. 
  Thus, as shown in cyan, the PN approximation can be applied when $\Lambda_c^{-1} \lesssim r_s$, where both finite-size and orbital effects are known or when $\Lambda_c^{-1} \lesssim r$, where orbital effects are known and finite-size effects are small enough to be neglected.
  \label{fig:softUV_sketch}}
\end{center}
  
\end{figure*}

There are several  phenomenological consequences associated to the new terms present in Eq.~(\ref{EFT_action}). Let us focus on binaries 
comprised of two BHs of similar size, whose Schwarzschild radii we collectively denote with $r_s$. If $\Lambda_c\gtrsim 1/r_s $, the  
observable deviations from GR present in our EFT mainly arise from modifications to the Kerr geometry of each individual BH, as well as, 
the quasi-normal modes (QNMs) of the remnant BH produced by their merger. Reference~\cite{Cardoso:2018ptl}  investigated how the GR Kerr geometry 
is modified in this regime, showing that: i) the equatorial frequency at the light ring and innermost stable circular orbit (ISCO),  
and the spin-induced quadrupole moment receive corrections, ii) the $\mathbb{Z}_2$ symmetry of the geometry is broken, iii) 
the tidal Love numbers become non-vanishing, and iv) the QNM spectrum is modified with respect to GR even 
for non-spinning BHs. Various thermodynamical properties of BHs in this EFT were also recently studied in Ref.~\cite{Reall:2019sah}.
We collectively refer to the corrections to binary dynamics and GW signal corresponding to these deformations of the Kerr geometry in \texttt{EFTGR} as finite-size effects; in  Appendix~\ref{app:finitesize} we show that these finite-size effects are too weak to be observable with current GW events, 
and therefore we are not yet able to put any constraints in the regime $\Lambda_c\gtrsim 1/r_s$.

Therefore in this paper we will instead focus on the opposite regime
$\Lambda_c\lesssim 1/r_s $, whose phenomenological consequences were
studied in Ref.~\cite{Endlich:2017tqa} in larger detail (in
particular, see Sec.~7 of Ref.~\cite{Endlich:2017tqa} for the
discussion of the two regimes).  In this case, the EFT breaks down at
distances larger than the gravitational radius and, due to the
assumption of soft UV completion, the finite-size effects
saturate~\footnote{The application of the assumption of a soft UV
    completion to this case goes as follows: one can think of a BH as
    a system with the characteristic scale $E\sim1/r_s$, then, for
    $r_s>\Lambda_c^{-1}$, finite-size effects enter the predictions
    for a binary system with the scaling $v^n/(r_s\Lambda_c)^m$ for
    some positive powers $n, m$. If not for the assumption,
    extrapolating this scaling naively, for $r_s<\Lambda_c^{-1}$, one could
    expect very large finite size effects, however, the assumption
    states that for $r_s<\Lambda_c^{-1}$ one can replace all the factors of
    $1/(r_s\Lambda_c)$ with some number not larger than order
    one. Thus, modifications to the tidal Love numbers and
    spin-induced multipole moments of the BH geometry need not
    dramatically affect the evolution of a binary system.}, and
  the dominant signature of {\tt EFTGR} in the GW signal arises from
  modifications of the two-body interactions between the BHs. As
  we will discuss in more detail below, these are corrections that
  enter at 2PN order, though they are additionally suppressed by a
  factor of $1/(\Lambda_c r)^6\ll1$, with $r$ being the separation of
  the two BHs.  Explicitly, the relative correction to the amplitude
and phase of the GWs scales as $v^4 /(\Lambda_c r)^6\sim v^{16}
/(\Lambda_c r_s)^6$, which can also be thought of as an 8PN term,
albeit with an extraordinarily large coefficient.
The relative size of finite-size effects and these new orbital effects---as well as the regimes of applicability of perturbative methods through which they are computed---are depicted schematically in Fig.~\ref{fig:softUV_sketch}; this figure also illustrates the saturation of corrections beyond the cutoff scale of our EFT, as imposed by the assumption of a soft UV completion.

In this regime (i.e., when $\Lambda_c\lesssim 1/r_s $), the corrections 
due to the modification to the intrinsic Kerr geometry cannot be computed 
explicitly, as this would require the full UV completion of the EFT.  
One can however estimate them at the level of order of magnitude. Finite-size effects from modifications to the spin-induced and tidally-induced
  quadrupole moments produce corrections to binary motion (and the
  subsequent GW signal) that scale as $\chi^2 v^4$ and $v^{10}$,
  respectively, where $\chi$ is the dimensionless spin of the BHs. 
Note that we have used our assumption of a soft UV completion to
  eliminate the scaling with $\Lambda_c r_s$ in these terms; as
  detailed in the footnote above, we have assumed that these effects
  saturate for BHs of size $r_s \sim \Lambda_c^{-1}$, and thus this
  scaling can simply be replaced by an order one number in the regime
  we work $r_s \lesssim \Lambda_c^{-1}$. Using this
  simplification, we estimate the relative impact of these finite-size
  effects as compared to the orbital effects discussed above by
  computing the total number of GW cycles each contribute to the
  inspiral. For BHs of mass greater than $\sim 10 M_\odot$, we find
  that the orbital effects typically contribute at least 100 times as
  many cycles as finite-size effects over the frequency bandwidth of
  the Advanced LIGO and Virgo detectors. Thus, we can safely
  neglect finite-size effects in our current study, leaving these
  details for future work.

\section{Gravitational-wave prediction for the inspiral regime in the effective--field-theory extension of 
General Relativity}
\label{sec:waves}

In this section we explicitly show that because {\tt EFTGR} modifies the conservative 
and dissipative dynamics of a two-body system, it leads to new terms in the GW 
phasing, which might be observed or constrained by GW observations. 

In the early-inspiral phase the dynamics of a binary system of two objects with mass $m_1$
and $m_2$ can be characterized by an effective action that in center-of-mass frame takes the
form~\cite{Goldberger:2004jt,Endlich:2017tqa}:
\begin{equation}\label{eff_action}
S = \int dt \left [ m_1 +m_2 +\frac{1}{2} \mu \mathbf{v}^2(t) - V(r(t)) + \frac{1}{2} Q_{ij}(t) R^{i0j0} +\ldots \right ],
\end{equation}
where $\mu=m_1m_2/(m_1+m_2)$ is the reduced mass of the system, $\mathbf{v}(t)$ is the relative velocity between the objects, $ V(r(t))$ is the potential energy, $Q_{ij}(t)$  is the mass quadrupole moment of the system and $\ldots$ denote higher-order multipole moments.

As shown in Ref.~\cite{Endlich:2017tqa}, in the PN and
$\Lambda_c\lesssim 1/r_s $ regimes, the leading-order correction to
the gravitational potential of a binary system with non-spinning or
slowly-spinning components is generated by the term $\mathcal{C}^2$ in Eq.~\eqref{EFT_action}, 
and reads:
\begin{equation}\label{potential}
V_{\Lambda}=512\,\frac{m_1 m_2}{r}\left(\frac{1}{\Lambda r}\right)^6\frac{(m_1^2+m_2^2)}{r^2}\,.
\end{equation}
The new terms in the action~\eqref{EFT_action} also modify the
gravitational radiation emitted by the binary. The leading order
correction comes again from the term  $\mathcal{C}^2$~\cite{Endlich:2017tqa}, and 
it can be written as a renormalization of the Newtonian quadrupole-moment $Q^{\rm Newt}_{ij}$
of the binary system:
\begin{eqnarray}\label{quadrupole}
Q_{ij} &=& Q_{ij}^{\rm Newt} + \cdots + Q_{ij}^\Lambda \,,\\
Q^\Lambda_{ij} &=& 2688\,\left(\frac{1}{\Lambda r}\right)^6\left(\frac{m_1+m_2}{r}\right)^2\,Q^{\rm Newt}_{ij}\,,
\end{eqnarray}
where $\ldots$ denote higher order PN corrections in the qudrupole moment. 
Henceforth, we focus solely on these leading-order EFT corrections, and compute their
effect on the conservative and dissipative dynamics. We then use the balance equation and calculate 
the leading-order correction to the GW phase.
 
\subsection{Conservative and dissipative dynamics}

On the orbital timescale, the various time-dependent
quantities in the action~\eqref{eff_action} can be treated as constant
and the radiative terms can be neglected.  Restricting to
quasi-circular orbits and varying the action ~\eqref{eff_action} with
respect to $\mathbf{r}(t)$, the equations of motion of the binary
system can be written as:
\begin{equation}\nonumber
\mu \Omega^2 r  = \frac{d V}{dr} = \frac{d V_\text{Newt}}{dr} + \frac{d V_\Lambda}{dr} + \ldots \,,
\end{equation}
where $\ldots$ denote higher PN corrections. The above equation can be inverted to get a relation between 
the orbital radius $r$ and the orbital frequency $\Omega$. To leading order in $1/\Lambda$ one gets
\begin{equation}
r = \frac{M}{\left (M \Omega\right)^{2/3}}\left[1 - \frac{1536\left(m_1^2 +m_2^2\right)}{M^8 \Lambda^6} \left(M \Omega \right)^{16/3}\right]\,.
\end{equation}

Using the equations above we find that the binding energy (per unit total mass) is given by
\begin{equation}
E = \frac{1}{2} \nu \mathbf{v}^2 + \frac{V(r)}{M} = -\frac{1}{2} \nu v^2 - 2\,560 \nu (1-2\nu) \left(\frac{d_\Lambda}{M}\right)^6 v^{18},
\end{equation}
where $M=m_1+m_2$ is the total mass of the system, $\nu=m_1m_2/M^2$ is the symmetric mass ratio, and we define $v \equiv (M \Omega)^{1/3}$. For convenience we also introduce the parameter $d_\Lambda \equiv 1/\Lambda$. Restoring the higher PN corrections in GR we have
\begin{equation}
E^{\Lambda}(v)=E^{\rm GR} - 2\,560\nu \,(1-2\nu)\left(\frac{d_{\Lambda}}{M}\right)^6  v^{18}\,.
\end{equation}
where $E^{\rm GR}$ denotes the PN expression for the binding energy in GR.

Furthermore, the renormalized quadrupole moment~\eqref{quadrupole} leads to corrections to the GW flux. 
As in GR, Eq.~\eqref{eff_action} predicts a leading-order GW flux given by the quadrupole formula
\begin{equation}
\mathcal{F} = \frac{1}{5} \langle \dddot{Q}_{ij} \dddot{Q}^{ij}\rangle\,,
\end{equation}
where $\langle\cdots\rangle$ indicates the time-average over an orbit. 
The resulting GW flux can be written as
\begin{equation}
\mathcal{F}(v)=\mathcal{F}^{\rm GR}(v)+\mathcal{F}^{\Lambda}(v)\,,
\end{equation}
where $\mathcal{F}^{\rm GR}(v)$ is the PN expression for the flux in GR, and the leading-order correction 
to the flux $\mathcal{F}^{\Lambda}(v)$ reads
\begin{equation}
\mathcal{F}^{\Lambda}(v)= \frac{\nu^2}{5} \left(- 24\,576 +393\,216 \nu \right)\left(\frac{d_{\Lambda} }{M}\right)^6 v^{26}\,.
\end{equation}

\subsection{Gravitational waveform in the stationary phase approximation}

We are now in the position of computing the leading-order correction to
the GW phase in {\tt EFTGR}. 

In the quasi-circular, adiabatic inspiraling stage, 
it is common to compute the PN approximation of the GW signal 
in the Fourier domain, using the stationary phase approximation (SPA) 
(see, e.g., Refs.~\cite{Damour:1997ub,Buonanno:2009zt}). In this approximation
the frequency-domain waveform can be written as~\cite{Buonanno:2009zt}
\begin{eqnarray}\label{waveform}
\tilde{h}_{\rm SPA}(f)&=&\frac{A(t_f)}{\sqrt{\dot{F}(t_f)}}e^{i[\Psi_{\rm SPA}(t_f)-\pi/4]}\,,\nonumber\\
\Psi(t)&\equiv& 2\pi f t - \phi(t)\,,
\end{eqnarray}
where $\phi(t)$ is the orbital phase and $\pi F(t)= d\phi(t)/dt$ defines the instantaneous GW frequency $F(t)$.
The quantity $t_f$ is the saddle point where $d\Psi(t)/dt=0$ (i.e., the time at which the instantaneous GW 
frequency $F(t)$ is equal to the Fourier variable $f$). In the adiabatic approximation $\Psi$ and $t_f$ are given by~\cite{Buonanno:2009zt}
\begin{eqnarray}
t_f&=&t_{\rm ref} +M\int_{v_f}^{v_{\rm ref}}\frac{E'(v)}{\mathcal{F}(v)} dv\,,\\
\Psi_{\rm SPA}(t_f)&=& 2\pi f t_{\rm ref} - \phi_{\rm ref}+2\int_{v_f}^{v_{\rm ref}}(v_f^3-v^3)\frac{E'(v)}{\mathcal{F}(v)} dv\,,\nonumber \\\label{spa_phase}
\end{eqnarray}
where we define $v_f\equiv (\pi M f)^{1/3}$, while $t_{\rm ref}$ and $ \phi_{\rm ref}$ are integration constants, and $v_{\rm ref}$ is an arbitrary reference velocity, commonly taken to be the velocity at the ISCO. 

Using the PN-expanded binding energy and flux, and expanding the ratio $E'(v)/\mathcal{F}(v)$  at the consistent PN order, the integral in Eq.~\eqref{spa_phase} can be solved explicitly. We find that the leading-order correction to the GW phase is given by
\begin{equation}\label{GWphase}
\Psi_{\rm SPA}(f) = \Psi^\text{GR}_{\rm SPA}(f) + \frac{3}{128 \nu v_f^5}\left(  \frac{234\,240}{11} - \frac{522\,240}{11} \nu  \right) \left(\frac{d_\Lambda}{M}\right)^6 v_f^{16}\,,
\end{equation}
where we $\Psi^{\rm GR}(f)$ represents the GW phase in GR~\footnote{Note that $v_{\rm ref}$ only contributes to terms that can be reabsorbed in the integration constants $t_{\rm ref}$ and $ \phi_{\rm ref}$, so the {\tt EFTGR} correction does not depend on the choice of $v_{\rm ref}$.}. 

As already anticipated above, an important conclusion of this calculation is that although the {\tt EFTGR} corrections are formally at 8PN order, 
when $d_{\Lambda}v^2/M \equiv (\Lambda r)^{-1} \sim 1$ the corrections are numerically at 2PN order. 

Lastly, we restrict the amplitude of the {\tt EFTGR} SPA waveform in Eq.~(\ref{waveform}) to 
the GR expression, because we expect that non-GR corrections to the phase dominate the signal.

\subsection{On the validity of the waveform model in the effective--field-theory extension of General Relativity}

As already mentioned the {\tt EFTGR} described by the action~\eqref{EFT_action}, and hence the SPA waveform ~\eqref{GWphase}, loses 
validity for orbital separations $r \lesssim d_{\Lambda}$. Therefore, when employing the {\tt EFTGR} SPA waveform,  
we should keep in mind that it is a reliable characterization of the gravitational radiation from the binary system 
\textit{only} for orbital scales larger than $d_{\Lambda}$. 

When comparing the signal with the data it is therefore convenient to
translate the orbital separation $r$ into a characteristic frequency
in the data such that we can identify the frequencies at which our
waveform model can be trusted.  There is no unique way to do this since $r$ is not a gauge-invariant quantity. A simple choice is to 
employ the Kepler's law and relate the orbital (angular) frequency 
to the orbital separation as $r = (M/ \Omega^2)^{1/3} + \mathcal{O}(v^2/c^2)$. 
Since, for quasi-circular orbits, the GW frequency is given by 
twice the orbital frequency we define the quantity
\begin{equation}\label{flambda}
f_{\Lambda}\equiv \frac{1}{\pi}\,\sqrt{\frac{M}{d_{\Lambda}^3}}\,,
\end{equation}
such that the {\tt EFTGR} SPA waveform given by Eqs.~\eqref{waveform} and ~\eqref{GWphase} 
remain valid for GW frequencies $f\ll f_{\Lambda}$. We emphasize that the equation above is not fundamental (e.g., we could have added PN corrections to it), and the total mass parameter $M$ plays the role of a typical mass scale that allows us to relate $d_{\Lambda}$ to a frequency scale.

In addition, when using a PN waveform model we also need to restrict to 
velocities such that $v\ll 1$.  There is no unique
choice for when the PN expansion is no longer valid, but a typical choice is
to use the frequency of the ISCO, $f_{\rm ISCO}$, as the cutoff. In the 
test-particle limit for a Schwarschild BH this is given by $f_{\rm ISCO}={1}/(6\sqrt{6}\pi M)$, 
where $M$ is the total mass of the system. We therefore assume the model to be only valid for
frequencies $f<f_{\rm cutoff}= \min[f_{\rm ISCO},f_{\rm high}]$, where
$f_{\rm high}<f_\Lambda$ is the cutoff frequency up to which we trust {\tt EFTGR}. 

Finally, there is no obvious prescription for the choice of $f_{\rm
      high}$, but generically an EFT can be employed without significant
    modifications until higher-order corrections become comparable to
    the leading one. The calculation of the corrections in {\tt EFTGR} 
    that are subleading in $1/(\Lambda r)$ goes beyond the scope of the
    present paper. However, we can estimate the frequency at which the
    {\tt EFTGR} breaks down by comparing the leading correction to some quantity to the GR contribution. As we have already discussed, our
    expansion contains, in fact, two small parameters: $1/(\Lambda r)$ [or $(d_\Lambda/r)$] and $v(r)$. 
    
From Eq.~(\ref{quadrupole}), we see that the leading-$(d_{\Lambda}/r)$ correction to the quadrupole goes as $\delta Q_{d_\Lambda}/Q_{\rm Newt}\sim
      (d_\Lambda/r)^6 v(r)^4$, thus the subleading-$(d_\Lambda/r)$ correction would be, 
      schematically, 
      of order $\delta Q^{\rm sub}_{d_\Lambda}/Q_{\rm Newt}\sim (d_\Lambda/r)^8 v^4(r)$,
      where, again, we highlighted only the dependence on the
      parameters and neglected numerical coefficients. We wish to
      estimate the radial separation $r_{\rm low}$, or equivalently the frequency
      $f_{\rm high}$, at which $\delta Q^{\rm sub}_{d_\Lambda}/Q_{\rm Newt}\sim \delta
      Q_{d_\Lambda}/Q_{\rm Newt}$. Clearly, we are going to obtain $r_{\rm low}\sim
      1/d_\Lambda$, but the numerical factors are quite important due
      the high-exponent in $r$ by which our effect
      scales. Furthermore, since we do not compute explicitly the
      subleading correction, we need to estimate this by looking only
      at our leading answer. Since the terms we consider depends both
      on $d_\Lambda/r$ and on $v^4(r)$, we can estimate when the
      expansion in $d_\Lambda/r$ breaks down only if we also have an
      estimate of when the $v$ expansion does. In fact, if we do this
      estimate when the $v$ expansion breaks down, our resulting
      estimate for $f_{\rm high}$ is expected to be independent of the
      PN order of the expression we use for the estimate (as it
      should).  Therefore, since we expect the PN expansion to break
      down at the ISCO orbit, we estimate that the $d_\Lambda/r$
      expansion breaks down when $\delta Q_{d_\Lambda}/Q_{\rm
        Newt} \sim 1$ [or $(d V_{\Lambda}/dr)/(d V_{\rm Newt}/dr) \sim 1$], with $v$ evaluated at $f=f_{\rm ISCO}$. 
      Let us notice that, given our choice of $f_{\rm high}$, at the maximum frequency we analyze the signal (i.e., 
$f_{\rm ISCO}$), the {\tt EFTGR} correction is at most comparable to GR. At lower frequencies, the $d_\Lambda$ corrections decrease more 
rapidly than GR. So the modifications are smaller than or equal to the GR contribution at all frequencies we analyze. 
Finally, for equal-mass binaries, the two conditions from the
      quadrupole and the gradient of the potential lead to almost
      identical estimates: $f_{\rm high}\approx 0.35
      f_\Lambda$.    
      
We emphasize that these estimates are only indicative of the frequency above which we expect the EFT to break down. In particular, our choice for the definition of  $f_{\Lambda}$ [see Eq.~\eqref{flambda}] is certainly not unique and does not arise as a  fundamental prediction of the theory. For example, one could have added PN corrections to Eq.~\eqref{flambda}, in which case $f_{\rm high}$ would also depend on the BH spins and on the mass ratio.  Ultimately, the question of what is the appropriate $f_{\rm high}$ can only be addressed by explicitly computing the next-order corrections, which we leave to future work.   

Therefore, to account for the fact that the current criteria are clearly not sharp, though are expected to be reasonably accurate, we use several values of $f_{\rm high}$ nearby $0.35 f_\Lambda$ to account for the sensitivity of our analysis to this choice.

\section{Gravitational constraints using Bayesian-selection methods}\label{sec:results}

\subsection{Setting bounds on gravity theories constructed  within the effective--field-theory approach}

To constrain the parameter $d_{\Lambda}$ we use a Bayesian model-selection approach, as we now briefly review. 

Consider a model or gravity theory $\mathcal{H}_i$ that we wish to test in light of the observed data $d$. Bayes' theorem 
tells us that the posterior probability density of the model given the data can be computed through:
\begin{equation}
P(\mathcal{H}_i|d,I)=\frac{P(\mathcal{H}_i|I)P(d|\mathcal{H}_i,I)}{P(d|I)}\,,
\end{equation}
where $I$ denotes all the prior information that one holds,
$P(\mathcal{H}_i|I)$ is the prior probability of the model
$\mathcal{H}_i$, $P(d|I)$ is a normalization constant and
$P(d|\mathcal{H}_i,I)$ is the marginal likelihood, also known as the
evidence, for the model $\mathcal{H}_i$.  If the model $\mathcal{H}_i$
is described by a set of parameters $\boldsymbol{\theta}= \{\theta_1,
\theta_2, \dots \}$, the marginalized likelihood
$P(d|\mathcal{H}_i,I)$ is given by
\begin{equation}
P(d|\mathcal{H}_i,I)=\int d\boldsymbol{\theta} P(\boldsymbol{\theta}|\mathcal{H}_i,I)P(d|\boldsymbol{\theta},\mathcal{H}_i,I)\,,
\end{equation}
where $P(\boldsymbol{\theta}|\mathcal{H}_i,I)$ is the prior
probability of the parameters $\boldsymbol{\theta}$ within the model
$\mathcal{H}_i$ and $P(d|\boldsymbol{\theta},\mathcal{H}_i,I)$ is the
likelihood of the observed data $d$ assuming a given value of the
parameters $\boldsymbol{\theta}$ in the model $\mathcal{H}_i$.

Consider now two competing theories $\mathcal{H}_i$ and $\mathcal{H}_j$ that we wish to compare given the observed data. To quantify the strength of one model against the other in describing the data, one can compute the ratio of the posterior probabilities, also known as the odds ratio:
\begin{equation}
\mathcal{O}_{i}^{j} = 
\frac{P(\mathcal{H}_{j} |d,I)}{P(\mathcal{H}_{i} |d,I)}=
\frac{P(\mathcal{H}_{j}|I)}{P(\mathcal{H}_{i}|I)}\frac{P(d|\mathcal{H}_j,I)}{P(d|\mathcal{H}_i,I)}=
\frac{P(\mathcal{H}_{j}|I)}{P(\mathcal{H}_{i}|I)}B_{i}^{j}\,,
\end{equation}
where in the last step we define the Bayes factor $B_{i}^{j}=P(d|\mathcal{H}_j,I)/P(d|\mathcal{H}_i,I)$, and the ratio $P(\mathcal{H}_{j}|I)/P(\mathcal{H}_{i}|I)$ denotes the prior odds.
A large Bayes factor $B_{i}^{j}\gg 1$ indicates that the data favor model $\mathcal{H}_{j}$ over $\mathcal{H}_{i}$, while a small Bayes factor $B_{i}^{j}\ll 1$ implies that $\mathcal{H}_{i}$ is preferred over $\mathcal{H}_{j}$. 
Here we are interested in comparing a GR waveform model against the {\tt EFTGR} waveform model described in the previous section.

The framework presented above can also be used to take advantage of multiple detections in order to increase our confidence for 
or against a given model. Consider a set of $N$ independent GW events with independent data sets given 
by $\mathbf{d}=\{d_1,d_2,\ldots,d_N\}$. We can write a combined odds ratio for the catalog of GW events as~\cite{Li:2011cg}:
\begin{equation}
\mathcal{O}_{i}^{j} = 
\frac{P(\mathcal{H}_{j} |\mathbf{d},I)}{P(\mathcal{H}_{i} |\mathbf{d},I)}=
\frac{P(\mathcal{H}_{j}|I)}{P(\mathcal{H}_{i}|I)}\frac{P(\mathbf{d}|\mathcal{H}_j,I)}{P(\mathbf{d}|\mathcal{H}_i,I)}=
\frac{P(\mathcal{H}_{j}|I)}{P(\mathcal{H}_{i}|I)}\, ^{(C)}B_{i}^{j}\,,
\end{equation}
where we define the combined Bayes factor as $^{(C)}B_{i}^{j}=P(\mathbf{d}|\mathcal{H}_j,I)/P(\mathbf{d}|\mathcal{H}_i,I)$.
This can be further simplified by noting that for a set of $N$ independent events with unrelated parameters $\boldsymbol{\theta}$ one has~\cite{Li:2011cg,Zimmerman:2019wzo}
\begin{equation}
P(\mathbf{d}|\mathcal{H}_i,I)=\prod_{k=1}^{N} P(d_k| \mathcal{H}_i,I)\,,
\end{equation}
and therefore 
\begin{equation}\label{comb_BF}
^{(C)}B_{i}^{j}=\prod_{k=1}^{N}\, ^{(k)}B_{i}^{j}\,,
\end{equation}
where $^{(k)}B_{i}^{j}$ denotes the Bayes factor for the event $k$.

\begin{figure*}[htb]
\begin{center}
\includegraphics[width=0.42\textwidth]{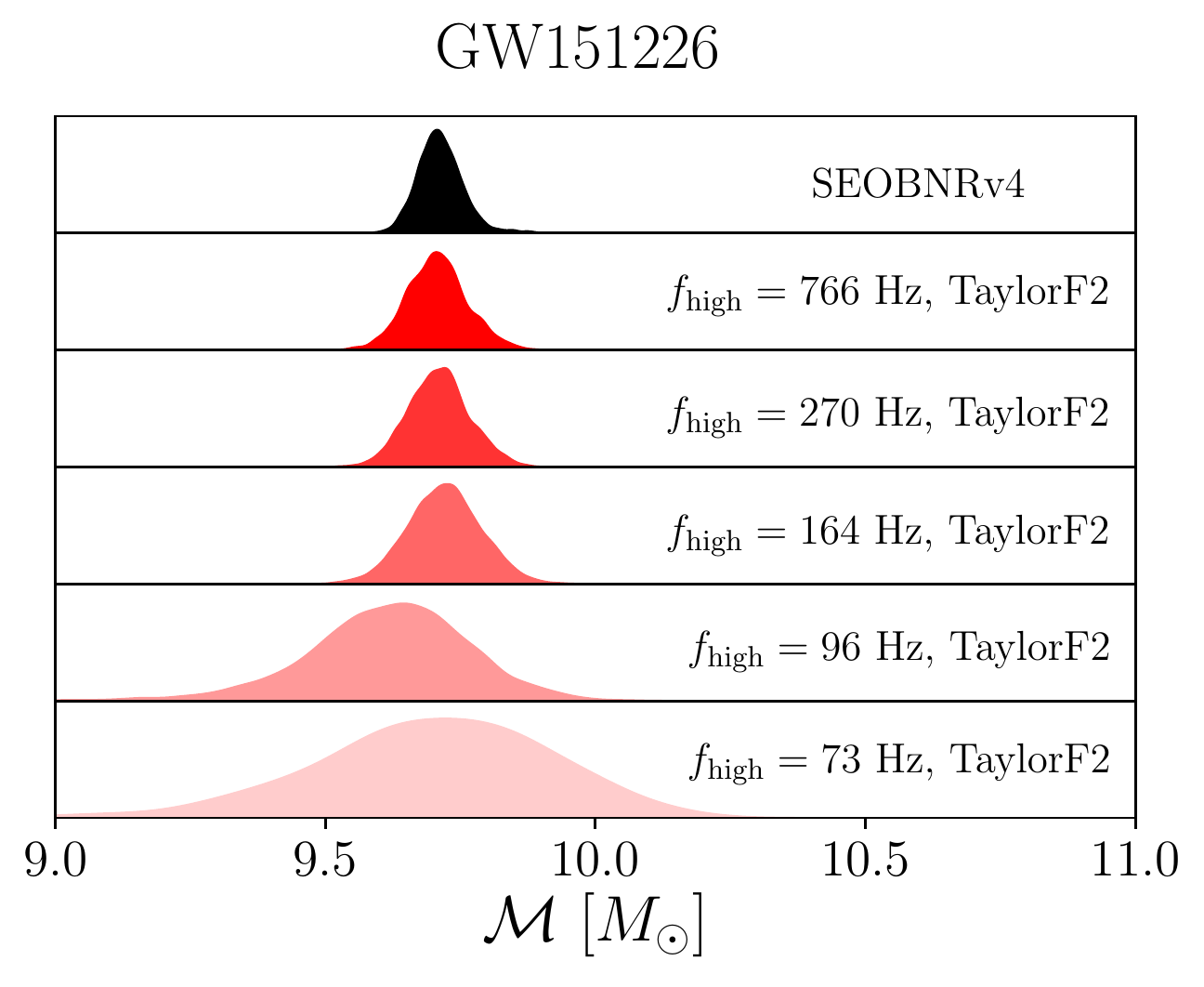} \hspace{0.8cm}
\includegraphics[width=0.42\textwidth]{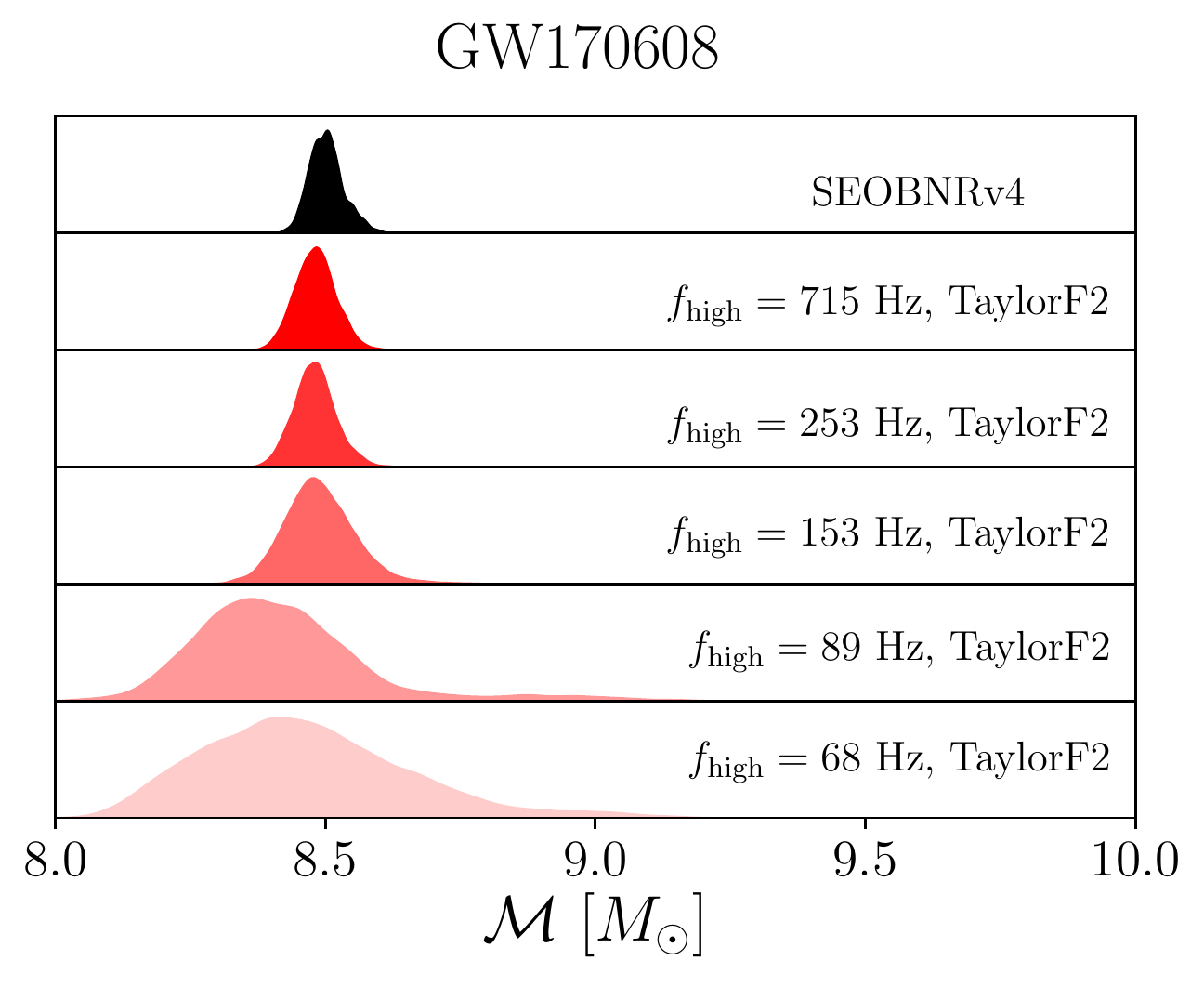}
\caption{Comparison between the posterior density distributions of the detector's frame chirp mass when using an IMR approximant (SEOBNRv4) and a inspiral-only PN approximant (TaylorF2) with different values of $f_{\rm high}$. For both GW151226 (left panel) and GW170608  (right panel) the results when using the PN approximant are in good agreement with the IMR result. \label{fig:m_chirp}}
\end{center}
\end{figure*}

Finally, to make inferences on the unknown parameters $\boldsymbol{\theta}$ for a given event, we use again Bayes' 
theorem to compute the posterior probability
\begin{equation}\label{posterior}
P(\boldsymbol{\theta}|d,\mathcal{H},I)=\frac{P(\boldsymbol{\theta}|\mathcal{H},I)P(d|\boldsymbol{\theta},\mathcal{H},I)}{P(d|\mathcal{H},I)}\,.
\end{equation}
For the prior probability density $P(\boldsymbol{\theta}|\mathcal{H},I)$ we follow the choices in Ref.~\cite{Veitch:2014wba}.
The likelihood function $P(d|\boldsymbol{\theta},\mathcal{H},I)$ is defined as the distribution of the residuals, assuming they are distributed as Gaussian noise colored by the noise-power spectral-density $S_n(f)$ for each detector~\cite{Veitch:2014wba}:
\begin{equation}
\label{eq:lik}
P(d|\boldsymbol{\theta},\mathcal{H},I) \propto \exp\left[-\frac{1}{2} \sum_{k} \left\langle h_k(\boldsymbol{\theta}) - d_k \middle| h_k(\boldsymbol{\theta}) - d_k\right\rangle\right],
\end{equation}
where the sum in the above expression is over all the detectors and we define the noise-weighted inner product as~
\cite{Finn:1992wt}:
\begin{equation}
\label{eq:overlap}
\left\langle g| h\right\rangle = 2\int_{f_{\rm low}}^{f_{{\rm cutoff}}} \frac{\tilde{g}(f)\tilde{h}^*(f)+\tilde{g^*}(f)\tilde{h}(f)}{S_n(f)}\,,
\end{equation}
where $\tilde{g}$ and $\tilde{h}$ denote the Fourier transform of $g$ and $h$, respectively.
To sample the posterior density functions and compute the evidences we use a nested sampling algorithm as implemented in the LALInference code~\cite{Veitch:2014wba} of the LIGO Algorithm Library Suite (LALSuite)~\cite{lalsuite}. 

For the minimum frequency in the analysis we use $f_{\rm low}=20$ Hz.
On the other hand, as we already discussed, when setting $f_{\rm
  high}$ we must take into account the fact that {\tt EFTGR} is
expected to break down when $f>f_{\Lambda}$, where we remind that
$f_{\Lambda}$ is defined through Eq.~\eqref{flambda}.  Therefore when
computing the likelihood function we set $f_{\rm high}< f_{\Lambda}$.
When choosing the value of $f_{\rm high}$, one must also take into
account the fact that the precise value of the GW frequency $f$ at
which the model breaks down is unknown. The choice of $f_{\rm high}$ therefore adds a systematic error in the constraints on $d_{\Lambda}$. 
To study the impact of this choice we work with several values 
of $f_{\rm high}$. It is clear that the smaller the ratio $f_{\rm high}/f_\Lambda$ is, the more conservative is the bound we place on the {\tt EFTGR} 
parameter $\Lambda$.

Assuming that one has a precise measurement of the total mass, for our lowest choice, $f_{\rm high}\approx 0.2f_\Lambda$, the effects on the physical observables from the {\tt EFTGR} 
corrections are smaller than the GR values for all the data points we use. 
Moreover, $f_{\rm high}\approx 0.35f_\Lambda$ allows the {\tt EFTGR} effects to be comparable to the GR effect for the highest frequencies included in the analysis, consequently, one may expect significant corrections from the subleading {\tt EFTGR} terms in this case. 
 
 To constrain $d_{\Lambda}$ we perform a Bayesian
  model selection where we compare, on a given data subset 
  where the EFT analysis can be reliably performed, the hypothesis
  that the data contains a signal described by the {\tt EFTGR}
  model~\eqref{GWphase} with a \emph{given} value of $d_{\Lambda}$
  against the hypothesis that the signal is described by GR, which
  corresponds to $d_{\Lambda}=0$. To fix $f_{\rm high}$ we also need to define the mass scale appearing in Eq.~\eqref{flambda}. Here we choose this mass scale to be given by the median source's frame mass obtained when using a full inspiral-merger-ringdown (IMR) template in GR. This choice is
  not unique, however we note that any other choice for this mass
  scale can be absorbed into a rescaling of $f_{\rm high}$. By considering
  a range of values for $f_{\rm high}$ we can therefore also
  incorporate this uncertainty in our results, as we discuss in more
  detail below.

\begin{figure*}[htb]
\begin{center}
\includegraphics[width=0.48\textwidth]{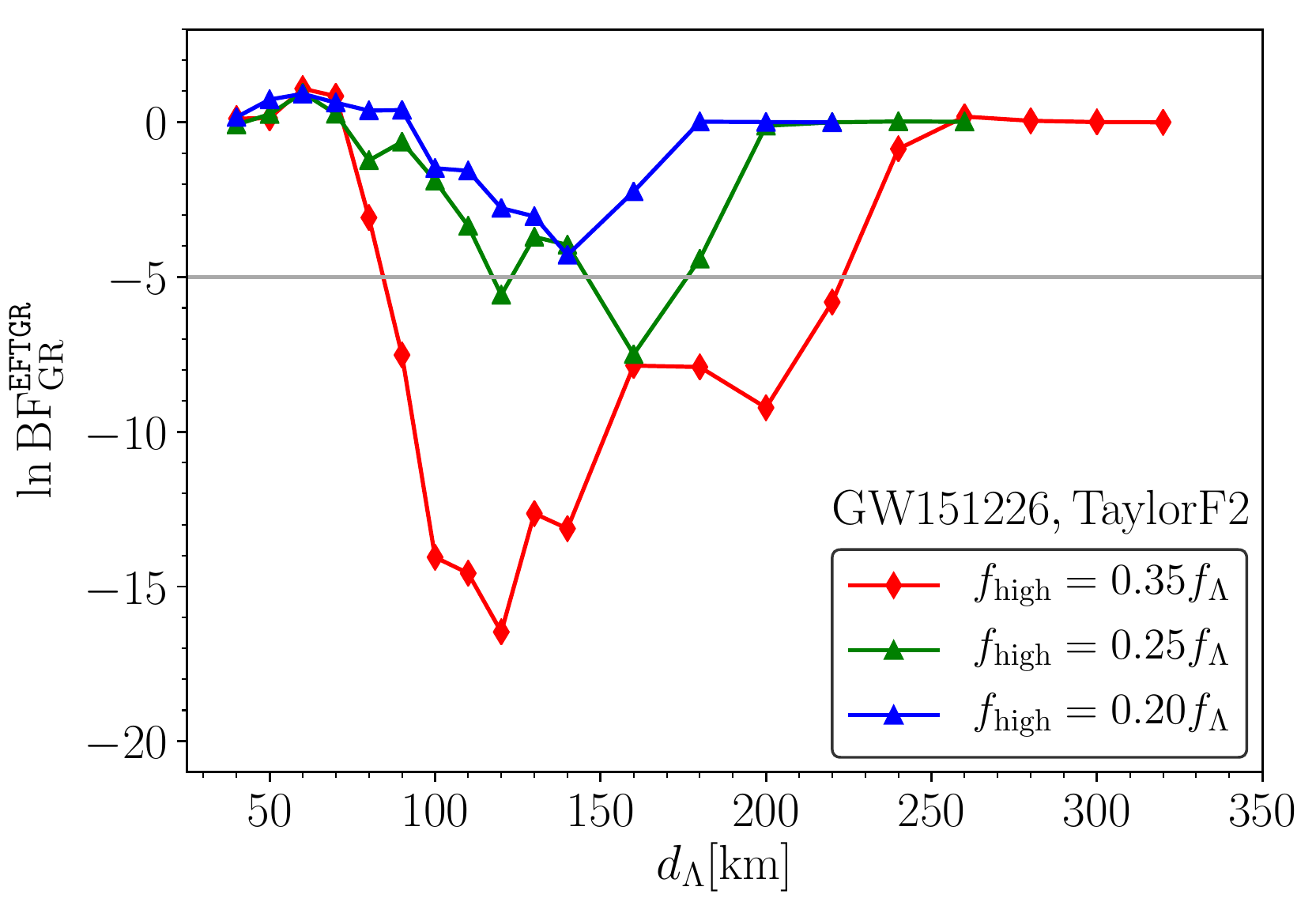} \hspace{0.5cm}
\includegraphics[width=0.48\textwidth]{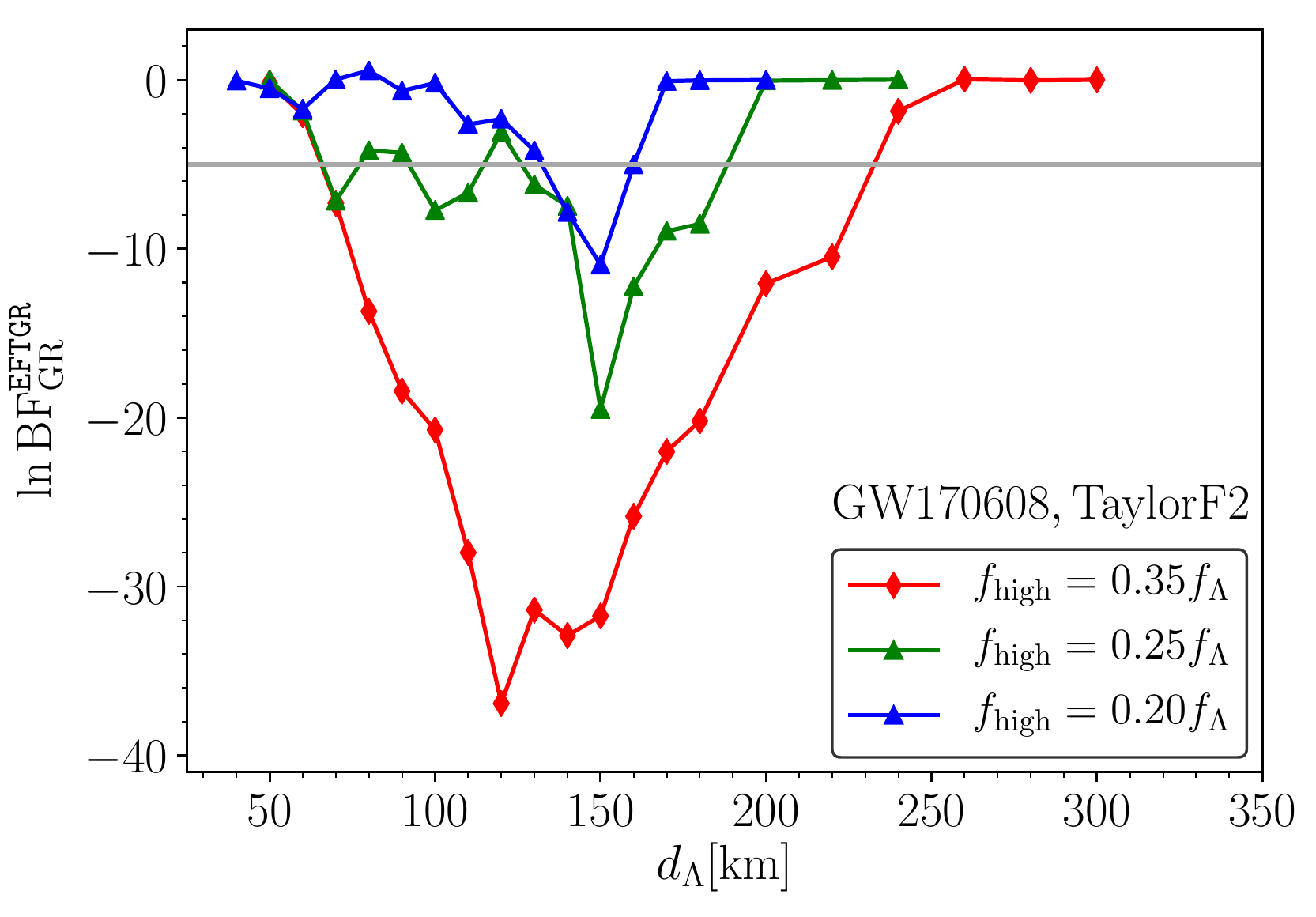}
\caption{The natural logarithm of the Bayes factor of the {\tt EFTGR} versus GR waveform model for different choices of $d_{\Lambda}$ (in km)
when using the inspiral-only PN waveform model. 
We use different values for $f_{\rm high}$ when computing the likelihood function~\eqref{eq:lik} to account for the different systematic uncertainties mentioned in the text.  
To compute $f_{\Lambda}$ we use the median source's frame total masses obtained with IMR templates as reported in Refs.~\cite{Abbott:2016nmj, Abbott:2017gyy}.
We show the results for the two lowest-mass GW events detected so far: 
GW151226 with total mass $M_{\rm tot}\simeq 22 M_{\odot}$ (left panel) and GW170608 with total mass $M_{\rm tot}\simeq 19 M_{\odot}$  
(right panel). For reference we mark the threshold $\ln B_{\rm GR}^{\rm {\tt EFTGR}}= -5$ with a grey solid curve. 
\label{fig:BF_TaylorF2}}
\end{center}
\end{figure*}
  In addition to these difficulties, the {\tt EFTGR} corrections to
  the GW phase given by Eq.~\eqref{GWphase} are only valid in the PN-expanded approximation 
of the inspiral phase, and do not provide any information about the expected
  non-GR corrections in the merger-ringdown phase of the signal.
  Therefore to constrain $d_{\Lambda}$ we employ the Flexible Theory Agnostic (FTA) code 
developed for LIGO and Virgo analyses~\cite{Abbott:2018lct,LIGOScientific:2019fpa}, and add the non-GR correction in
  Eq.~\eqref{GWphase} to an aligned-spin GR inspiral-only PN waveform model valid up
  to 3.5PN order.\footnote{In the LIGO Algorithm Library~\cite{lalsuite}
    the specific name of the waveform models that we use in this paper
    are TaylorF2 (see, e.g., Ref.~\cite{Buonanno:2009zt}),
    IMRPhenomD~\cite{Khan:2015jqa} and
        SEOBNRv4~\cite{Bohe:2016gbl}} This waveform is artificially set to zero for
  frequencies $f>f_{\rm ISCO}$, the regime at which the PN expansion
  is expected to loose significant accuracy.  We should emphasize that choosing
  $f_{\rm ISCO}$ as the frequency cutoff of our waveform model is
  largely arbitrary, and therefore this specific choice adds another
  systematic effect in the constraints that we can place on
  $d_{\Lambda}$.  To quantify this uncertainty in
  Appendix~\ref{app:IMR} we also show our results when using an IMR 
waveform, finding negligible differences. 

One might worry that the use of an inspiral-only PN approximant containing an abrupt
cutoff at $f_{\rm ISCO}$ might lead to a bias in the measured
parameters if part of the post-inspiral signal is loud enough, even
when assuming GR~\cite{Mandel:2014tca}.  We have indeed checked that
for GW150914~\cite{Abbott:2016blz}, a GW event with a significant
signal-to-noise ratio (SNR) in the merger-rigdown part of the signal,
using a GR inspiral-only PN template leads to biases in the measured BH masses when
compared to the measurement done with an IMR model.  However, for
low-mass GW events that, in the frequency band of the detectors, are
mostly dominated by the inspiral part of the signal, we do not expect
this to be a problem.  To justify this statement we have checked that for
the two lowest-mass binary BH (BBH) events reported thus far by the LIGO and Virgo
collaborations, GW151226~\cite{Abbott:2016nmj} and
GW170608~\cite{Abbott:2017gyy}, we indeed find no biases in the
measurement of the waveform parameters when using a GR inspiral-only PN waveform
model~\footnote{GW151226 and GW170608 were detected with a network SNR
  of $\sim 12$ and $\sim 15$, and with total source frame masses of
  $M_{\rm tot}\simeq 22 M_{\odot}$ and $M_{\rm tot}\simeq 19
  M_{\odot}$, respectively.}.  This can be seen in
Fig.~\ref{fig:m_chirp} where we show the marginalized posterior
density distributions for the chirp mass that we obtain when using the
inspiral-only PN-approximant TaylorF2 (with $d_{\Lambda}=0$) for different values of
$f_{\rm high}$, and compare these results with the posterior obtained
when using an IMR template (SEOBNRv4) with $f_{\rm high}=1\,024$ Hz.
As expected, the statistical error increases with decreasing $f_{\rm
  high}$, due to the decreasing SNR, but the measurements when using
the IMR template and the inspiral-only PN template are in good agreement. 
The same conclusion holds for the other parameters of the model. The loss of SNR when $f_{\rm high}\lesssim 200$ Hz is consistent with the fact that for both events $f_{\rm ISCO}\sim 200$ Hz, and therefore, part of the signal is cut whenever $f_{\rm high}\lesssim 200$ Hz. This feature will be important to understand the qualitative behaviour of our results in the next Section. Given these
conclusions in what follows we focus our tests on the two longest signals 
detected so far by LIGO and Virgo, notably GW151226 and
GW170608.

\subsection{Bounds on the effective--field-theory extension of General Relativity  using GW151226 and GW170608}

Using the Bayesian framework presented above we now show that GW151226
and GW170608 can already be used to set meaningful constraints on the parameter
$d_{\Lambda}$. 

Our main results are summarized in Fig.~\ref{fig:BF_TaylorF2} where we 
show the natural logarithm of the Bayes factor of the {\tt EFTGR} hypothesis against 
GR, $\ln B_{\rm GR}^{\rm {\tt EFTGR}}$. We remind that the {\tt EFTGR} hypothesis corresponds to the
hypothesis that the data are well described by the
{\tt EFTGR} waveform model~\eqref{GWphase} with a \emph{given} value of $d_{\Lambda}$
against the hypothesis that the signal is described by GR, i.e.,
$d_{\Lambda}=0$.  We show our results when using an inspiral-only PN approximant in
Fig.~\ref{fig:BF_TaylorF2}, but similar results hold when using an IMR
waveform (see Appendix~\ref{app:IMR} for more details). To quantify
the systematic error in the uncertainty of the frequency at
which the {\tt EFTGR} model cannot be trusted, we consider different choices of $f_{\rm
  high}=[0.2,0.25,0.35]f_{\Lambda}$. As mentioned above, to compute $f_{\Lambda}$ we fix the mass scale in Eq.~\eqref{flambda} to be the median value for the total BH mass obtained when estimating the parameters with an IMR waveform model in GR as given in Refs.~\cite{Abbott:2016nmj, Abbott:2017gyy}. We could have made other choices for this mass scale, however  the uncertainty in the total mass of the system can be absorbed inside the definition of $f_{\rm high}$. For example, using the 90\% credible intervals reported in Refs.~\cite{Abbott:2016nmj, Abbott:2017gyy} we have verified that this would be equivalent to rescaling $f_{\rm high}$ by a factor $\sim 0.96$ and $\sim 1.12$ for the lower and upper limits of the 90\% credible intervals, respectively. Taking $f_{\rm high}=0.25 f_{\Lambda}$ as our fiducial choice, this variation is smaller than the difference between the three different values of $f_{\rm high}$ that we consider.
Therefore by using several values for $f_{\rm high}$ when showing our results we are also taking into account this uncertainty.

\begin{figure}[htb]
\begin{center}
\includegraphics[width=0.48\textwidth]{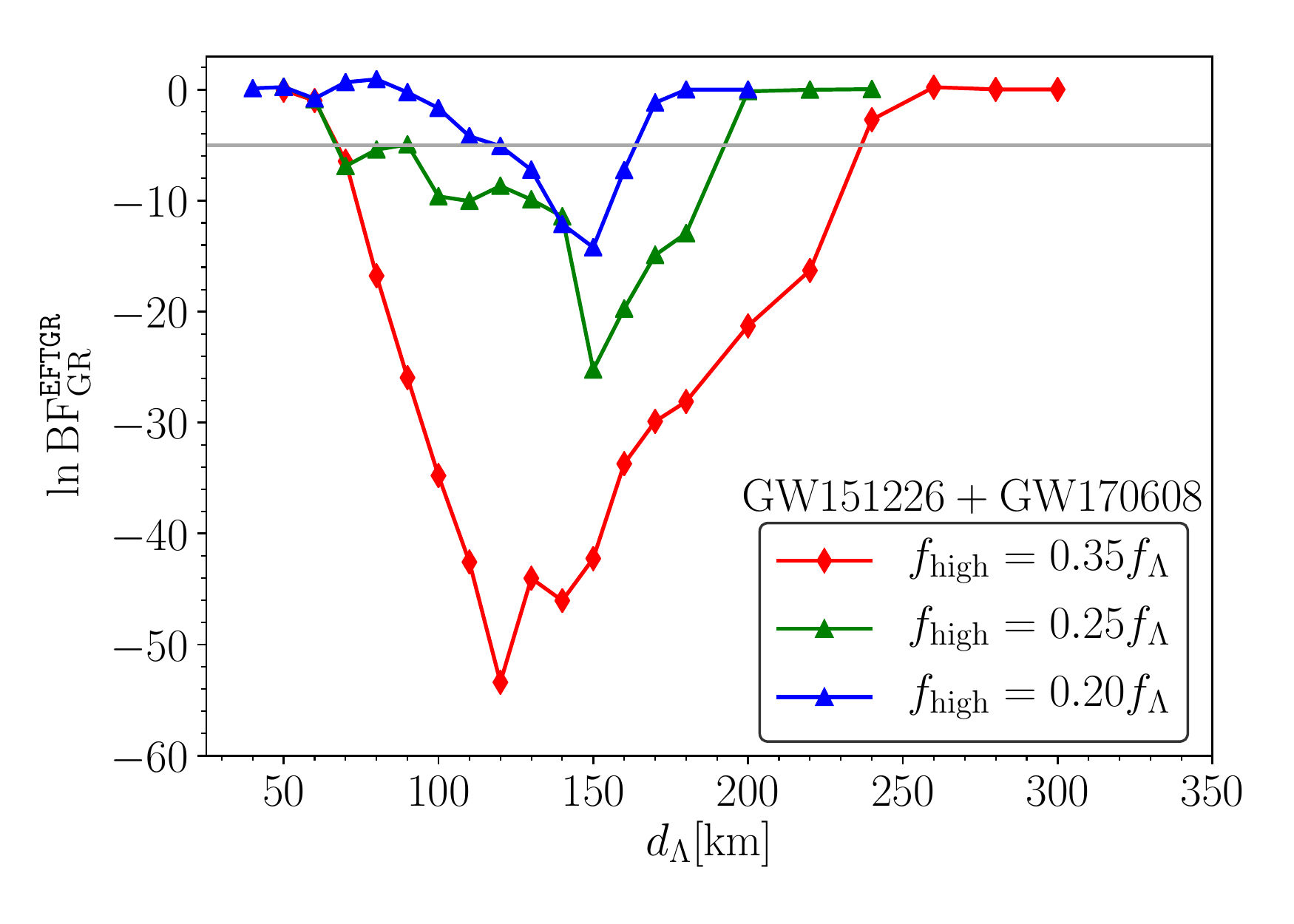}
\caption{The log Bayes factors of the {\tt EFTGR} versus GR waveform when combining the results shown in Fig.~\ref{fig:BF_TaylorF2}.
  \label{fig:BF_combined}}
\end{center}
\end{figure}

Our results show that $\ln B_{\rm GR}^{\rm {\tt EFTGR}}< 0$ for a wide
range of values for $d_{\rm \Lambda}$, implying that the {\tt EFTGR}
hypothesis is disfavored for some values of $d_{\rm \Lambda}$. The
constraints we obtain are similar between the two events, due to their
similar total mass. However, for GW170608 one obtains slightly stronger
evidence against the {\tt EFTGR} hypothesis due to the larger SNR of this
event when compared with GW151226. Setting a threshold $\ln B_{\rm
  GR}^{\rm {\tt EFTGR}}\lesssim -5$~\footnote{In the typically used
  classification of Ref.~\cite{jeffreybook}, $B_{\rm GR}^{\rm
    {\tt EFTGR}}<1/100$ corresponds to decisive evidence for GR against the
  {\tt EFTGR} hypothesis.} to constrain $d_{\Lambda}$, we can draw the
strong conclusion that values of $d_{\Lambda}$ between roughly $70$
and $200$ km are strongly disfavored by the data, independently on our
choice of $f_{\rm high}$. The evidence against the {\tt EFTGR} hypothesis
is, however, highly dependent on $f_{\rm high}$. This should be
expected given that the effect of $d_{\Lambda}$ only appears in the
waveform phase~\eqref{GWphase} at high PN order and therefore, for a
given $d_{\Lambda}$, the larger $f_{\rm high}$ the larger the
contribution from the non-GR term and the stronger the constraints
that we can place.

The qualitative features of our results can be largely understood in
terms of two competing effects: (1) for small $d_{\Lambda}$ the Bayes
factor asymptotically goes to $B_{\rm GR}^{\rm {\tt EFTGR}}\sim 1$, since
the non-GR contribution decreases as $d_{\Lambda}$ becomes smaller,
and (2) in the opposite direction, for increasing values of
$d_{\Lambda}$, one must also decrease $f_{\rm high}$ and therefore for
some large value of $d_{\Lambda}$ the amount of data that we use are no
longer sufficient to dig out the signal from the noise, no matter
which model we use. Therefore, for some large value of $d_{\Lambda}$
the Bayes factor should also approach $B_{\rm GR}^{\rm {\tt EFTGR}}\sim 1$,
as we indeed see in Fig.~\ref{fig:BF_TaylorF2}. Due to these two
competing effects, we are only able to constrain an interval of values
$d^{\rm low}_{\Lambda}<d_{\Lambda}<d^{\rm high}_{\Lambda}$.  The
lower-limit $d^{\rm low}_{\Lambda}$ in this constraint is more robust
against different choices of $f_{\rm high}$, since for large enough
$f_{\rm high}$ one is effectively analysing the full signal up to the
PN-aproximant frequency cutoff, $f_{\rm ISCO}$.  On the other hand,
the upper-limit $d^{\rm high}_{\Lambda}$ is obviously quite dependent
on the choice that we make for $f_{\rm high}$, since $d^{\rm
  high}_{\Lambda}$ mostly depends on the maximum value of $f_{\rm
  high}$ below which the matched-filtered SNR is too small to separate
the signal from the noise.

Using Eq.~\eqref{comb_BF}, the results presented in
Fig.~\ref{fig:BF_TaylorF2} can be combined to improve the constraints
on $d_{\Lambda}$. The resulting combined Bayes factors are shown in
Fig.~\ref{fig:BF_combined}. One can see that by combining both events
the evidence against the {\tt EFTGR} hypothesis around
$d_{\Lambda}\sim 150$ km increases significantly when compared to the
constraints obtained with a single event. The importance of
  combining the information from different GW events is even more
  striking for the case $f_{\rm high}=0.2 f_{\Lambda}$. For that
  choice of $f_{\rm high}$, GW151226 alone does not give any
  meaningful constraints, however when combining it with GW170608, one
  gets strong constraints for values of $d_{\Lambda}$ around
  $d_{\Lambda}\sim 150$ km. We expect these constraints to further
improve as more GW events are observed in the future.

% ----------------------------------------------------------------------------------- % 
\begin{table}
  \caption{Range of values of $d_{\Lambda}$ that are strongly disfavored by GW151226 and GW170608, and when combining both events. 
To determine these constraints we set the threshold at $\log B_{\rm GR}^{\rm {\tt EFTGR}}\lesssim -5$. We show 
the constraints for the different choices of $f_{\rm high}$.}
\label{tab:BFs}
\centering
\begin{tabular}{c@{\quad} c@{\quad}c@{\quad}c@{\quad} c}
\toprule
\multirow{2}{*}{Event}    &  \multicolumn{3}{c}{$d_{\Lambda}$ [km]}  \\
\cline{2-4}
        &       $f_{\rm high}=0.2 f_{\Lambda}$  &  $f_{\rm high}=0.25 f_{\Lambda}$ & $f_{\rm high}=0.35 f_{\Lambda}$ \\
\colrule
GW151226 	& 	 --  & 	$\sim  [125,175]$ &   $\sim [85,225]$\\	
GW170608	& 	$\sim [135,160]$  & 	$\sim [65,190]$ &   $\sim [65,230]$ \\	
Combined  	& 	 $\sim [120,165]$  & 	$\sim [65,190]$ &   $\sim [65,235]$  \\
\botrule
\end{tabular}
\end{table}
% ----------------------------------------------------------------------------------- % 

The constraints of the individual and the combined results are
summarized in Table~\ref{tab:BFs}. 
Taking into account the systematics entering 
the choice of $f_{\rm high}$, we conclude that coupling constants around $d_{\Lambda}\sim  150$ km are strongly disfavored. This is in agreement with the
prediction made in Ref.~\cite{Endlich:2017tqa}, and can be easily
understood from the fact that in this range $d_{\Lambda}/r_{\rm
  ISCO}\sim \mathcal{O}(1)$ for the BBHs we considered. This regime is
the region where the deviations from GR are maximized while still
keeping the perturbative control of the EFT. Interestingly, this sets
a typical scale that a given event can constrain. Combining the
information from more events will therefore not only be important to
increase the confidence for or against a given value of $d_{\Lambda}$,
but also to increase the range of values of $d_{\Lambda}$ that one can
probe and possibly constrain.

Finally, so far we have assumed that $\Lambda^6>0$, but as mentioned in
Sec.~\ref{sec:EFT}, there is no strong reason to dismiss the case
$\Lambda^6<0$. Thus, we have also extended the analysis discussed in this
section to the case $\Lambda^6<0$~\footnote{Note that for
  $\Lambda^6<0$ we can define $d_{\Lambda}\equiv 1/|\Lambda|$ such
  that the non-GR corrections can be easily obtained by doing the
  transformation $d_{\Lambda}^6 \to -\,d_{\Lambda}^6$ in
  Eq.~\eqref{waveform}.}, and have found that when combining the results
from both events, the constraints become $1/|\Lambda| \sim
[140,160]$ km, $1/|\Lambda| \sim [65,190]$ km and $1/|\Lambda| \sim
[65,235]$ km for $f_{\rm high}=[0.2,0.25,0.35]f_{\Lambda}$,
respectively.

\section{Conclusions}\label{sec:conc}

We presented the first constraints on the {\tt EFTGR} proposed in Ref.~\cite{Endlich:2017tqa}, which is 
described by the action~\eqref{EFT_action}, using the two lowest-mass 
GW events so far detected by LIGO and Virgo. 
Taking into account the different uncertainties in our analysis, our results show that current observations disfavour coupling constants of 
the order $d_{\Lambda}\equiv 1/\Lambda \sim 150$ km. The Bayesian
method we employed can be easily used to combine information from 
several observations, and therefore we expect that future GW detections 
will allow us to further constrain this theory.

Our constraints assumed that, in the regime where $\Lambda\lesssim
1/r_s$, the leading order non-GR effect come from the properties of
the binary and not modifications from the BH geometry.  As argued in
Sec.~\ref{sec:EFT}, the situation may be different for highly
spinning BHs in which case, within the soft-UV completion assumption,
modifications to the spin-induced BH quadrupole moment may
dominate. By adding order one corrections to the GR spin-induced BH
quadrupole moment, we have checked that for the events we considered
the modifications to the binary dynamics always dominates. This is in
agreement with recent work, where it was shown that the spin-induced
BH quadrupole moments cannot be well constrained for GW151226 and
GW170608~\cite{Krishnendu:2019tjp}. Constraining the spin-induced BH
quadrupole moment with future GW events is an interesting direction
that may further improve our constraints.

We focused our analysis to the regime $\Lambda\lesssim
1/r_s$. Complementary constraints in the regime $\Lambda\gtrsim
1/r_s$, can in principle be obtained from, e.g., measurements of the
spin-induced BH quadrupole moment, the tidal Love numbers or
measurements of the BH-remnant's QNMs~\cite{Cardoso:2018ptl}. The events 
observed by LIGO and Virgo so far have too low SNR to put any 
constraints in this regime (see Appendix~\ref{app:finitesize}). 
In fact, it would be interesting to study how 
well future detectors on the ground and in space, such 
as Cosmic Explorer, Einstein Telescope and LISA, can 
constrain $\Lambda$, not only because BBHs will have much higher SNR, but also 
much longer inspiral phase, thus making our constraints 
more robust by choosing a lower value of  $f_{\rm high}$.

Although we only showed constraints for $\Lambda$, the same method can
be applied to constrain $\tilde{\Lambda}$ and $\Lambda_-$, but 
constraints on the latter will likely require events with a
larger SNR and binaries with spinning components for which the effect 
of the terms $\tilde{\mathcal{C}}^2$ and
$\tilde{\mathcal{C}}\mathcal{C}$ is expected to be
non-negligible~\cite{Endlich:2017tqa}.  It would also be
  interesting to analyse {\tt EFTGR} with cubic terms in the Riemann
  tensor, though they are theoretically disfavored. 

  Finally, our results relied on computing the leading-order non-GR PN
  correction, but stronger constraints could in principle be obtained
  by building IMR waveforms for these EFTs. This will require
  numerical simulations of BBHs in the {\tt EFTGR}. Numerical
  simulations could be done along the lines of the proposals in
  Refs.~\cite{Okounkova:2017yby,Witek:2018dmd,Okounkova:2018pql,Okounkova:2019dfo}
  where significant progress has been made to perform numerical
  simulations of BH mergers in an EFT extension of gravity with
  higher-derivative operators and an additional light degree of
  freedom.

\acknowledgments

We thank Lijing Shao for his preliminary work in estimating the
detectability of the phenomena arising in the {\tt EFTGR}. V.G. and
L.S. thank Junwu Huang for discussions. 

R.B. acknowledges financial support from the European Union's Horizon
2020 research and innovation programme under the Marie Sk\l
odowska-Curie grant agreement No. 792862. V.G. is a Marvin L. Goldberger Member at IAS. V.G. and L.S. are partially supported by
the Simons Foundation Origins of the Universe program (Modern
Inflationary Cosmology collaboration). L.S. is partially supported by
NSF award 1720397. This research was supported by the Munich Institute for Astro- and Particle Physics (MIAPP) which is funded by the Deutsche Forschungsgemeinschaft (DFG, German Research Foundation) under Germany's Excellence Strategy - EXC-2094 - 390783311.

This research has made use of
data, software and/or web tools obtained from the Gravitational Wave
Open Science Center (https://www.gw-openscience.org), a service of
LIGO Laboratory, the LIGO Scientific Collaboration and the Virgo
Collaboration.  LIGO is funded by the U.S. National Science
Foundation. Virgo is funded by the French Centre National de Recherche
Scientifique (CNRS), the Italian Istituto Nazionale della Fisica
Nucleare (INFN) and the Dutch Nikhef, with contributions by Polish and
Hungarian institutes. 

The authors are grateful for computational resources provided by the LIGO Laboratory and supported by the National Science Foundation Grants PHY-0757058 and PHY-0823459.

\appendix

%%%%%%%%%%%%%%%%%%%%%%%%%%%%%%%%%%%%%%%%%
\begin{figure}[htb]
\begin{center}
\includegraphics[width=0.48\textwidth]{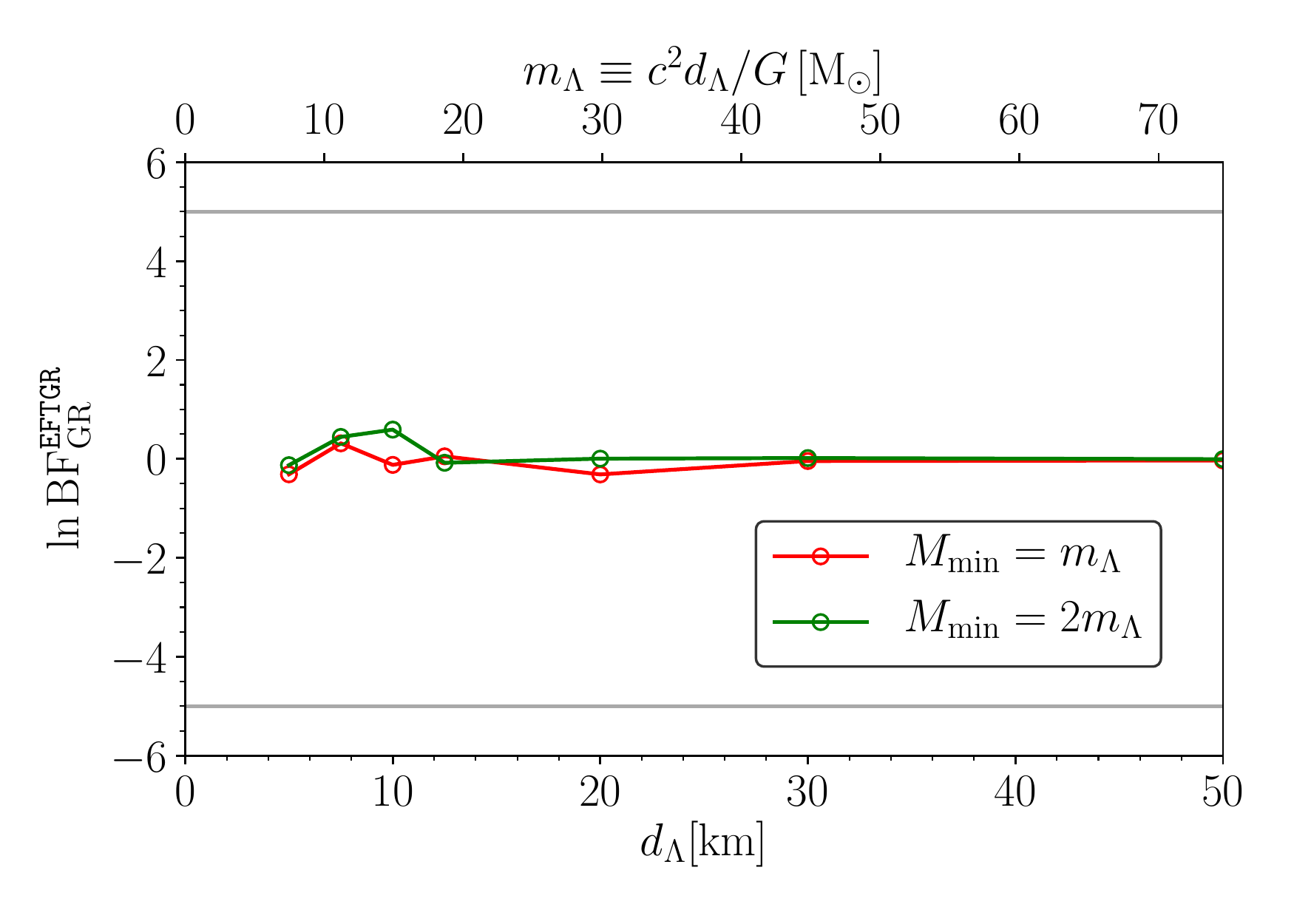}
\caption{The natural logarithm of the Bayes factors of the {\tt EFTGR} versus GR waveform for
  GW170608 in the regime where $d_{\Lambda}\lesssim M$, with $M$ the
  mass of either BH in the binary, with corresponding $m_{\Lambda}$
  shown in the top $x$-axis, when using the PN waveform model. We fix
  the minimum mass $M_{\rm min}$ in the priors for the BH masses such
  that $d_{\Lambda}< M$ is always satisfied in the GW template. To
  account for the systematic uncertainty in the validity of the
  waveform model, we use different values for the minimum mass in the
  prior $M_{\rm min}$ as shown by the different
  curves.\label{fig:finitesize}}
\end{center}
\end{figure}
%%%%%%%%%%%%%%%%%%%%%%%%%%%%%%%%%%%%%%%%%

\section{Finite-size effects for $\Lambda\gtrsim 1/r_s$}\label{app:finitesize}

%%%%%%%%%%%%%%%%%%%%%%%%%%%%%%%%%%%%%%%%%
\begin{figure}[htb]
\begin{center}
\includegraphics[width=0.48\textwidth]{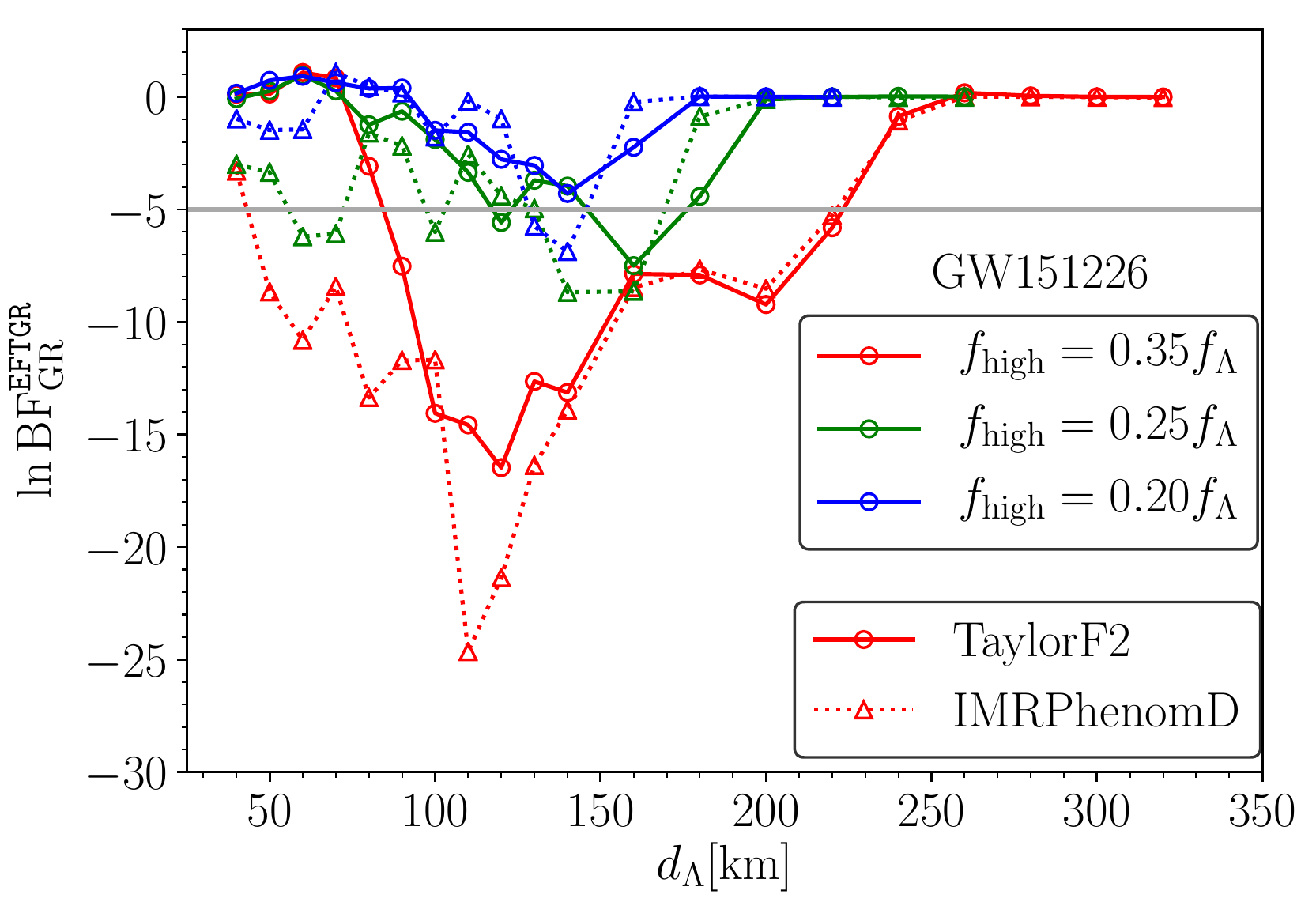}
\includegraphics[width=0.48\textwidth]{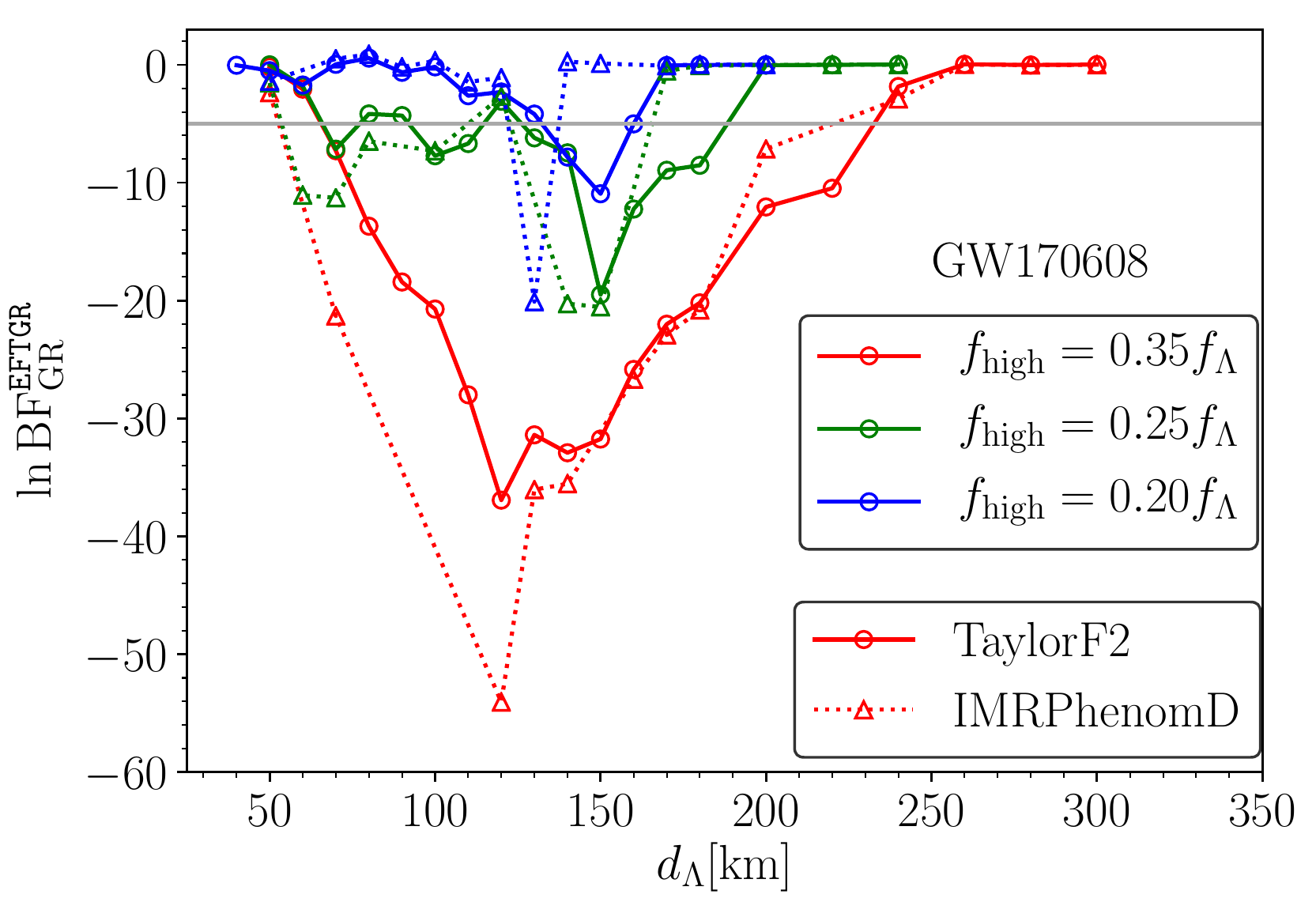}
\caption{Same as Fig.~\ref{fig:BF_TaylorF2}, but also showing the case where we use an IMR waveform model (IMRPhenomD). For the IMR model the {\tt EFTGR} correction is taped at the merger frequency that we take to be the peak of the waveform's amplitude\label{fig:BF_IMR}}
\end{center}.
\end{figure}
%%%%%%%%%%%%%%%%%%%%%%%%%%%%%%%%%%%%%%%%%

As explained in Sec.~\ref{sec:EFT}, in the regime where
$\Lambda\gtrsim 1/r_s$, or equivalently $d_{\Lambda}\lesssim M$, with
$M$ the mass of either BH in the binary, the modifications to the GW
template are mainly associated to modifications of the BH geometry.
For this case, the leading-order modifications in the inspiral phase 
come from a modified BH spin-induced quadrupole moment and
non-vanishing tidal Love number~\cite{Endlich:2017tqa,Cardoso:2018ptl}. 

Within the PN expansion, the influence of the spin-induced quadrupole moment enters the GW phase at
2PN order whereas the leading-order tidal Love number enters at 5PN
order (see, e.g., Ref.~\cite{Blanchet:2013haa}). By inserting the
explicit expressions for the BH spin-induced quadrupole moment and
tidal Love number found in Ref.~\cite{Cardoso:2018ptl} in a PN
template, we can therefore check whether we can impose any constraints
in this regime. For simplicity we only consider the $\Lambda$ term in
the action~\eqref{EFT_action}, but results would be similar for the
other terms. We also only consider GW170608 since this event has a
larger SNR than GW151226.

We used the same Bayesian model selection procedure presented in
Sec.~\ref{sec:results}, except that now no constraints need to be
imposed on the highest frequency in the data. Instead, since we are
considering the regime $d_{\Lambda}\lesssim M$, we define a mass
associated with $\Lambda$ given by $m_{\Lambda}\equiv \Lambda^{-1}$
and impose a minimum mass $M_{\rm min}$ in the prior probabilities of
the BH masses such that only masses satisfying $M_{\rm min}\geq
m_{\Lambda}$ are considered in the analysis. This ensures that
$d_{\Lambda}< M$ is always satisfied. Our results are shown in
Fig.~\ref{fig:finitesize}. They confirm that, independently on
$M_{\rm min}$, there is no preference for either GR or the {\tt EFTGR}
model. Therefore no constraints can be imposed in this regime.

\section{Constraints when using an inspiral-meger-ringdown waveform}\label{app:IMR}

In this appendix we show that similar constraints to the ones shown in
the main text can also be obtained with an IMR model. For the {\tt EFTGR}
waveform IMR model we include the non-GR correction in
Eq.~\eqref{GWphase} in the inspiral phase of the waveform and, since the corrections 
in the merger-ringdown are unknown, we taper the correction to zero at the merger 
frequency, taken to be the peak of the waveform's amplitude. We perform the test 
using the FTA code developed and employed in the analyses of 
Refs.~\cite{Abbott:2018lct,LIGOScientific:2019fpa}.

The results obtained when using an IMR model (IMRPhenomD) are shown in
Fig.~\ref{fig:BF_IMR}. For comparison we also show the results
obtained with an inpiral-only PN waveform (TaylorF2), already shown in
Fig.~\ref{fig:BF_TaylorF2}. We find that the range of values of
$d_{\Lambda}$ for which the GR waveform is preferred over the {\tt EFTGR}
one when using the IMR waveform is similar to the one obtained when
using an inspiral-only PN waveform (TaylorF2), the main difference being that for
$d_{\Lambda}\lesssim 150$ km the constraints obtained with the IMR
model are slightly stronger. This is no surprise given that this value
of $d_{\Lambda}$ is close to the approximate location of the final BH
ISCO for the BBH systems we consider. For $d_{\Lambda} \gtrsim
150$ km both the IMR and inspiral-only PN results should agree given that our choice
for $f_{\rm high}$ satisfies $f_{\rm high}\lesssim f_{\rm ISCO}$. On
the other hand, for $d_{\Lambda}\lesssim 150$ km the IMR waveform
model gives slightly stronger constraints due to the inclusion of
non-GR corrections at frequencies above $f_{\rm ISCO}$ in the
waveform. However, independently of whether we use the IMR or the inspiral-only 
PN approximant the range of values $d_{\Lambda}$ that we constrain does
not change significantly.

\bibliography{inspire}

\end{document}